\documentclass[twocolumn,amsmath,aps]{revtex4}

\usepackage{amssymb}
\usepackage{amsmath}
\usepackage{epsfig}
\usepackage{subfigure}
\usepackage{mathrsfs}
\usepackage{longtable}

\newcommand{\be}{\begin{equation}}
\newcommand{\ee}{\end{equation}}
\newcommand{\bea}{\begin{align}}
\newcommand{\eea}{\end{align}}

\newcommand{\unit}[1]{\ensuremath{\, \mathrm{#1}}}

\begin{document}

\title{Ultra-Light Scalar Fields and the Growth of Structure in the Universe}
\author{David J. E. Marsh}
\affiliation{Rudolf Peierls Centre for Theoretical Physics, University of Oxford, 1 Keble Road, 
Oxford, OX1 3NP, UK}
\author{Pedro G. Ferreira}
\affiliation{Astrophysics, University of Oxford, DWB, Keble Road, Oxford, OX1 3RH, UK}


\begin{abstract}
Ultra-light scalar fields, with masses of between $m=10^{-33}\unit{eV}$ and $m=10^{-22}\unit{eV}$, can affect the growth of structure in the Universe. We identify the different regimes in the evolution of ultra-light scalar fields, how they affect the expansion rate of the universe and how they affect the growth rate of cosmological perturbations. We find a number of interesting effects, discuss how they might arise in realistic scenarios of the early universe and comment on how they might be observed.
\end{abstract}

\maketitle
\section{Introduction}
\label{intro}

Ultra-light scalar fields arise generically in high energy physics, most commonly as axions or other axion-like particles (ALPs). They are the Pseudo-Goldstone bosons (PGBs) of spontaneously broken symmetries, which only acquire mass through non perturbative effects. In generic string theory compactifications we expect a number of axions~\cite{witten2006}, similar to the well known QCD axion~\cite{pecceiquinn1977, thooft1976a, thooft1976b, dine1981, preskill1983, steinhardt1983, turner1983, abbott1983, dine1983, turner1986}, and more recently e.g. \cite{visinelli2009}, and their cosmology has been well studied in the past (see, for example, \cite{banks1996}). However their symmetry breaking scale is much higher, $f_a \sim 10^{16}\unit{GeV}$, being set by the string scale and the instanton action. The axions arise from closed two cycles in the compact space: the symmetry breaking scale is only weakly dependent on the size of the cycle, so will be roughly constant across all the axions in a given compactification volume, whereas the mass of each axion depends exponentially on the size of the cycle and so we expect axions masses to evenly distribute on a logarithmic mass scale all the way down to the Hubble scale today, $H_0\sim 10^{-33}\unit{eV}$~\cite{axiverse2009}.

We can characterise the Lagrangian of such a generic ALP using two parameters: the symmetry breaking scale, $f_a$, and the overall scale of the potential, $\Lambda$. The axion field is an angular variable, $\theta$, since the path integral is unchanged by the shift symmetry $\theta \rightarrow \theta + 2\pi$. The effective four dimensional Lagrangian can be written as:
\begin{equation}
\mathcal{L} = \frac{f_a^2}{2}(\partial\theta)^2 - \Lambda^4 U(\theta)
\end{equation}
where $U(\theta)$ is some periodic potential. Bringing the kinetic term into canonical form we define the field $\phi = f_a \theta$, with Lagrangian:
\begin{equation}
\mathcal{L} = \frac{1}{2}(\partial \phi)^2 + V(\phi)
\label{eqn:lagrangian}
\end{equation}
where $V(\phi)$ is again a periodic potential. Expanding the potential in powers of $\phi/f_a$, all the couplings of the field $\phi$ come suppressed by the large scale $f_a$, and  to quadratic order we find that the mass is given by:
\begin{equation}
m^2 = \frac{\Lambda^4}{f_a^2}
\end{equation}
Hence, any axion is equivalently parameterised by its mass and symmetry breaking scale.

Production of cosmological axions proceeds by the vacuum realignment mechanism. When the Peccei-Quinn-like $U(1)$ symmetry is broken at the scale $f_a$ the axion acquires a vacuum expectation value, $\theta_i$, uncorrelated across different causal horizons. However provided inflation occurs after symmetry breaking, and with a reheat temperature $T\lesssim f_a$, then the field is homogenised over our entire causal volume. This is the scenario we consider in this paper. The field $\theta$ is a PGB and  evolves according to the potential $U(\theta)$ induced after symmetry breaking by instantons. Once the mass overcomes the Hubble drag the field begins to roll towards the minimum of the potential, in exact analogy to the minimum of the instanton potential restoring $\mathcal{CP}$ invariance in the Peccei-Quinn mechanism for the QCD axion. Coherent oscillations about this minimum lead to the production of the weakly coupled ALPs.~\cite{linde1991, hertzberg2008, sikivie2008}

In this paper, because of the weak couplings caused by the high scale $f_a$, we will choose to work simply with the quadratic part of the potential, and largely ignore the effects of anharmonicities, although we discuss them briefly in Section~\ref{Background}. In line with this choice we also make the generalisation to work with ultra-light scalar fields that do not respect the shift symmetry and therefore have no need for a periodic potential.

Scalar fields with masses in the range $10^{-33}\unit{eV}<m<10^{-22}\unit{eV}$ are also well motivated dark matter candidates and constitute what Hu has dubbed ``fuzzy cold dark matter'', or FCDM~\cite{hu2000}. The Compton wavelength of the particles associated to ultra-light scalar fields, in natural units, $\lambda_c = 1/m$, is of the size of galaxies or clusters of galaxies, an so the uncertainty principle prevents localisation of the particles on any smaller scale. This naturally suppresses formation of structure and serves as a simple solution to the problem of ``cuspy halos'', and the large number of dwarf galaxies, which are not observed and are otherwise expected in the standard $\Lambda$CDM cosmological model.

The large phase space density of ultra-light scalar fields causes them to form Bose-Einstein condensates (see~\cite{sikivie2009} and references therein) and allows them to be treated as classical fields in a cosmological setting. This could lead to many interesting, and potentially observable phenomena, such as formation of vortices in the condensate~\cite{silverman2002, kain2010}, and black hole super radiance~\cite{axiverse2009, arvanitaki2010,rosa2010} which could provide direct tests of the ``string axiverse'' scenario of~\cite{axiverse2009}. In this paper we will be concerned with the large scale effects of ultra-light scalar fields on structure formation by computing the resultant matter power spectrum in a cosmology where a fraction, $f=\Omega_a/\Omega_m$, of the CDM is made up of a such field.

If ALPs exist in the high energy completion of the standard model of particle physics, and are stable on cosmological time scales, then regardless of the specifics of the model Tegmark et al have argued~\cite{tegmark2006} that on general statistical grounds we indeed expect a scenario where they make up an order one fraction of the CDM, alongside the standard WIMP candidate of the lightest supersymmetric particle. However it must be noted that there are objections due to Mack and Steinhardt~\cite{mack2009a, mack2009b} when we consider a population of light fields in the context of inflation. The problem with these objections is that they make some assumptions about what we mean by ``fine tuning'' of fundamental physical theories, which is also related to the problem of finding a measure on the landscape of string theory and inflation models (see, for example,~\cite{linde2010}), the so called ``Goldilocks Problem''. Addressing these arguments in any detail is beyond the scope of this paper, but we consider the issue sufficiently unresolved, and ultra-light scalar fields to be sufficiently well motivated as dark matter candidates otherwise, to press on regardless in search of phenomenology.

In the context of generalized dark matter~\cite{hu1998b} we can see the effect of the Compton scale of these fields through the fluid dynamics of the classical field. The sound speed of a field with momentum $k$ and mass $m$ at a time where the scale factor of the FLRW metric is $a$ is given by:
\begin{align}
c_s^2 &= \frac{k^2}{4m^2 a^2}; \quad k<2ma \nonumber \\
c_s^2 &= 1; \quad k>2ma \nonumber \\
\label{eqn:cssquared}
\end{align}
On large scales the pressure becomes negligible, the sound speed goes to zero and the field behaves as ordinary dust CDM and will collapse under gravity to form structure. However on small scales, set by $\lambda_c$, the field becomes relativistic and the particles free-stream, suppressing the formation of structure. This observation is our main point of departure to consider the effect of ultra-light scalar fields on the matter power spectrum.

The paper is organised as follows: in Section \ref{Formalism} we set out our system of equations for the cosmological expansion and perturbations. In Section \ref{Background} we use analytic approximations and solutions to identify important scales in the evolution of the background field and discuss in more detail the issues concerning the fraction of dark matter from ALPs, and its production. We go on in Section~\ref{perturbations} to set up the initial conditions for perturbations in the metric, radiation, dust matter, and the scalar field. We then work analytically to approximate the effect of an ALP dark matter component on the matter power spectrum and see the emergence of the free-streaming scale in the problem. In Section \ref{Background_Results} we briefly present numerical results for the background evolution and the effect of ultra-light scalar fields on the cosmic expansion. Our main results on the effect of ultra-light scalar fields on the matter power spectrum are given in Section~\ref{Power_Spectrum}, where we give some useful parameterisations to describe them. Finally in Section \ref{Discussion} we discuss these effects and their possible detection and outline future directions of work. 

\section{The Formalism: Equations of Motion}
\label{Formalism}

We work in first order cosmological perturbation theory of the Friedmann-Lema\^{i}tre-Robertson-Walker (FLRW) metric, in the synchronous gauge, as presented in \cite{bertschinger1995}. The line element is:
\begin{equation}
ds^2 = a^2(\tau)[-d\tau^2 + (\delta_{ij}+h_{ij})dx^idx^j]
\end{equation}
where $a(\tau)$ is the scale factor, and $\tau$ is conformal time. The scalar modes of $h_{ij}$ can be written as a Fourier integral in terms of the two fields $h(\vec{k},\tau)$ and $\eta(\vec{k},\tau)$:
\begin{equation}
h_{ij}(\vec{x},\tau) = \int d^3k e^{i\vec{k}\dot\vec{x}}[\hat{\vec{k}}_i \hat{\vec{k}}_j h(\vec{k},\tau) + (\hat{\vec{k}}_i \hat{\vec{k}}_j - \frac{1}{3} \delta_{ij}) 6\eta(\vec{k},\tau)]
\end{equation}
where $\hat{\vec{k}}_i$ is a unit vector in the $i^{\mbox{th}}$ direction.

For a perfect fluid of energy density $\rho$ and pressure $P$ the energy momentum tensor is given by:
\begin{align}
T^0_{\ \ 0} &= -(\rho + \delta \rho) \nonumber \\
T^0_{\ \ i} &= (\rho + P)v_i \nonumber \\
T^i_{\ \ j}&= (P + \delta P)\delta^i_{\:j} \nonumber \\
\end{align}
where $\rho$ and $P$ are the average density and pressure, and $\delta \rho$ and $\delta P$ represent first order perturbations about homogeneity and isotropy. To zeroth order, the Einstein equations give the Friedmann equation:
\begin{equation}
\mathcal{H}^2 = \left(\frac{\dot{a}}{a}\right)^2 = \frac{8\pi G}{3}a^2\rho
\label{eqn:friedmann}
\end{equation}
where an overdot denotes a derivative with respect to conformal time $\tau$. The first order equations are:
\begin{align}
k^2 \eta - \frac{1}{2}\mathcal{H} \dot{h}&= \frac{1}{2}a^2 \delta T^0_{\ \ 0} \\
\ddot{h} +2 \mathcal{H} - 2k^2 \eta&=-a^2\delta T^i_{\ \ i}
\end{align}
from which $\eta$ can be eliminated, leaving us with a second order equation for $h$:
\begin{equation}
\ddot{h}+\mathcal{H}\dot{h} =a^2[\delta T^0_{\ \ 0} - \delta T^i_{\ \ i} ]
\label{eqn:h}
\end{equation}

To couple a scalar field to these equations we compute the energy momentum tensor from the potential in the usual way:
\begin{equation}
T^{\mu}_{\ \ \nu} = \phi^{;\mu}\phi_{;\nu} - {1\over2}(\phi^{;\alpha} \phi_{;\alpha} +2V)\delta^{\mu}_{\: \nu}
\end{equation}
Working to first order in perturbations about a homogeneous field:
\begin{equation}
\phi(\vec{k},\tau)=\phi_0(\tau) + \phi_1(\vec{k},\tau)
\end{equation}
we have, for a quadratic potential $V(\phi) = (1/2) m^2 \phi^2$:
\begin{align}
\rho_a = &\frac{a^{-2}}{2}\dot{\phi}_0^2 + \frac{m^2}{2}\phi_0^2 \label{eqn:rhoa} \\
\delta\rho_a =& a^{-2}\dot{\phi}_0\dot{\phi}_1 + m^2 \phi_0 \phi_1 \label{eqn:deltarho}\\
P_a =& \frac{a^{-2}}{2}\dot{\phi}_0^2 - \frac{m^2}{2}\phi_0^2 \label{eqn:pa}\\
\delta P_a =& a^{-2}\dot{\phi}_0 \dot{\phi}_1 - m^2\phi_0 \phi_1 \label{eqn:deltap} \\
(\rho + P)\theta_a =&a^{-2}k^2 \dot{\phi}_0 \phi_1  
\end{align}
The advantage of the synchronous gauge is that all of these quantities are independent of the metric perturbations. 

Next we require the equations of motion for $\phi_0$ and $\phi_1$, which are found from the Lagrangian, Eqn.~\ref{eqn:lagrangian} (equivalently we could use the conservation equations and Einstein equations to work directly with the fluid dynamical variables of Eqns.~\ref{eqn:rhoa}, \ref{eqn:deltarho}, \ref{eqn:pa}, \ref{eqn:deltap}~\cite{hu1998b}, but making this computationally tractable for scalar fields would require us to make further approximations that do not always hold in the regions of parameter space we are interested in):
\begin{align}
\ddot{\phi}_0 + 2\mathcal{H}\dot{\phi}_o + m^2 a^2 \phi_0 &= 0 \label{eqn:phi0} \\
\ddot{\phi}_1 +2\mathcal{H} \dot{\phi}_1 +(m^2 a^2 + k^2)\phi_1 &= -\frac{1}{2}\dot{\phi}_0\dot{h} \label{eqn:phi1}
\end{align}

To obtain the evolution equations for perturbations in the dust CDM and the radiation we use conservation of energy momentum, $T^{\mu\nu}_{\quad ;\mu} = 0$, in $k$-space to obtain the first order conservation equations:
\begin{align}
\dot{\delta}&=-(1+w) \left(  \theta+\frac{\dot{h}}{2} \right) - 3\mathcal{H}\left(  \frac{\delta P}{\delta\rho} -w \right)\delta \\
\dot{\theta}&=-\mathcal{H}(1-3w)\theta - \frac{\dot{w}}{1+w}\theta + \frac{\delta P/\delta\rho}{1+w}k^2 \delta - k^2\sigma
\end{align}
The variables $\theta$ and $\sigma$ are defined as:
\begin{align}
(\rho+ P)\theta &\equiv ik^j \delta T^0_{\ \ j} \\
(\rho+ P)\sigma &\equiv -(\hat{\vec{k}}_i \dot \hat{\vec{k}}_j - \frac{1}{3}\delta_{ij})\Sigma^{ij}
\end{align}
$\Sigma^i_{\ \ j}$ is the traceless component of $T^i_{\ \ j}$, a perturbation we henceforth ignore; $w=P/\rho$ is the equation of state, and $\delta = \delta\rho/\rho$ is the overdensity.

Working with no baryons coupled to the photon fluid, using the CDM particles as the comoving fluid that defines the synchronous gauge i.e. $\theta_c = 0$, and noting the equations of state  $w_{\gamma}=(P/\rho)_{\gamma}=(\delta P/\delta \rho)_{\gamma}=1/3$, and $w_c=(P/\rho)_c(\delta P/\delta\rho)_c=0$ we have:
\begin{align}
\dot{\delta}_c&=-\frac{1}{2}\dot{h} \label{eqn:deltam} \\
\dot{\delta}_{\gamma}&= -\frac{4}{3}\left(  \theta_{\gamma} + \frac{\dot{h}}{2} \right) \label{eqn:deltag}\\
\dot{\theta}_{\gamma}&=\frac{1}{4}k^2 \delta_{\gamma} \label{eqn:thetag}
\end{align}
Eqn.~\ref{eqn:deltam} can easily be integrated once we have the initial conditions to give $\delta_c = -\frac{1}{2} h$, so that the evolution of the matter becomes trivial and we need only work with $h$.

\section{Background Evolution and Production of ALPS in the Early Universe}
\label{Background}

We are interested in scenarios containing a fraction of the total energy density today in an ultra-light scalar field, therefore we would like to be able to specify $\Omega_a$ in terms of the initial displacement of the field, $\phi_i$, or equivalently the initial misalignment angle, $\theta_i$. To do this we look for an analytic solution to the equation of motion Eqn.~\ref{eqn:phi0}. This is most easily done in physical time, defined by $dt = a(\tau) d\tau$. In this subsection only, overdots will denote derivatives with respect to $t$, so that the Hubble parameter is given by $H(t)=\dot{a}/a$.

We work in reduced Planck units $1/m_{pl}^2=8\pi G=1$. We rescale to use dimensionless variables $t \rightarrow H_0 t$, $H\rightarrow H/H_0$, $\phi \rightarrow \phi/m_{pl}$, $m \rightarrow m/H_0$, where $H_0$ is Hubble today, and remain in these variables until we discuss the matter power spectrum in Section~\ref{Power_Spectrum}. The equations governing the background become:
\begin{equation}
\ddot{\phi}_0 + 3H\dot{\phi}_0 + m^2\phi_0 = 0 \label{eqn:phiphysical}
\end{equation}
\begin{equation}
H^2= \frac{\rho_a(t)}{3} + \frac{\Omega_c}{a^3}+ \frac{\Omega_{\gamma}}{a^4} +\Omega_{\Lambda}
\label{eqn:hubble} 
\end{equation}
where the density in ALPs is now given by:
\begin{equation}
\rho_a(t) = \frac{1}{2} \dot{\phi}_0^2 + \frac{1}{2} m^2 \phi_0^2
\label{eqn:rhohat}
\end{equation}

Eqn.~\ref{eqn:phiphysical} can be solved in terms of Bessel functions if we take the ansatz $a(t)\propto t^p$, which is true in both radiation dominated (early time), and matter dominated (late time) eras, giving:
\begin{equation}
\phi_0(t) = a(t)^{-3/2}(m t)^{1/2}(AJ_n(m t) + BY_n(m t))
\label{eqn:phibessel}
\end{equation}
with $n = (1/2)\sqrt{9p^2 - 6p + 1}$.  We ignore the $Y_n$ solution since it is singular at early times where we know that $\phi_0$ should take its value from the misalignment angle. The asymptotic forms of $J_n$ tell us how the energy density in a scalar field redshifts at early and late times and exhibits a well know feature of scalar field evolution in an expanding universe. For $m t \ll 1$:
\begin{equation}
J_n(m t) \approx \frac{1}{\Gamma (n+1)} \left (  \frac{m t}{2} \right )^n
\end{equation} 
Substituting into Eqn. \ref{eqn:phibessel}, along with $a\propto t^p$ yields:
\begin{equation}
\phi_0(t) \propto t^{-\frac{3}{2}p}t^{\frac{1}{2}}t^n
\end{equation}
which gives $\phi_0=\mathbf{const.}$ for both the radiation dominated era ($p=1/2$, $n=1/4$) and the matter dominated era ($p=2/3$, $n=1/2$). This in turn shows that the energy density remains a constant in this regime: at early times the energy density in ALPs redshifts like a cosmological constant.

Later, such that $m t \gg 1$, we have that:
\begin{equation}
J_n(m t) \approx \left ( \frac{2}{\pi m t} \right )^{1/2} \cos \left(  m t - \frac{n\pi}{2} - \frac{\pi}{4} \right )
\end{equation}
Substituting $H=p/t$ now gives:
\begin{align}
\Omega_a(t) &= \frac{A^2m^2}{3\pi}\frac{1}{a^3}\left(  1+ \frac{9}{4}\frac{p^2}{(m t)^2} \cos^2 \left( m t -\frac{n \pi}{2} - \frac{\pi}{4} \right) \right) \nonumber \\
& \propto  \frac{1}{a^3} + \mathcal{O}((m t)^{-2}) \nonumber \\
\label{eqn:laterho}
\end{align}
for all values of $p$. We see that at late times the energy density in axions redshifts like ordinary matter.

What these simple observations do not tell us about is the transition from cosmological constant ($\Lambda$) behaviour to Dark Matter (DM) behaviour, and how this transition can affect the expansion rate and age of the universe if it contains a significant fraction of DM in ALPs. As we will see later there are novel effects even here in the background.

For now we will continue to work analytically and delineate two important scales in the evolution, an important region of ALP parameter space, and set the initial condition on $\phi_0$ for a given $\Omega_a$.

The axion field starts oscillating in the crossover between the two asymptotic expressions for the Bessel function, when $m t_{osc}\approx 1$, which is the same order as the time when the mass overcomes the Hubble drag, $m \approx 3H(t_{osc})$. This defines one scale in the problem. The background evolution will depend on whether this occurs in the radiation or matter dominated era, defined by $\rho_m(a_{eq})=\rho_{\gamma}(a_{eq})$, where $\rho_m=\rho_c+\rho_a$, the total density in matter. If the field has begun oscillations in the radiation dominated era then it will be redshifting like matter and contribute as usual when deriving $a_{eq} \simeq \Omega_{\gamma}/\Omega_m$. However, if the ALPs begin oscillating in the matter dominated era, they will be redshifting as a cosmological constant at equality and will contribute negligibly to $\rho_m$. In particular, we can ignore $\rho_a(t)a^4/\rho_c(t_0)$, so that if the ALPs make up a fraction, $f$, of the total density in matter today, we obtain the modified formula for the scale factor at equality:
\begin{equation}
a_{eq} \simeq \frac{\Omega_{\gamma}}{\Omega_m}\frac{1}{(1-f)}
\end{equation}
For ALPs that begin oscillations in the matter era, this change to the redshift of equality will have knock-on effects for the estimation of other cosmological parameters and could possibly place tight constraints on such a light species making up an order one fraction of the total dark matter density.

The temperature of the CMB fixes $\Omega_\gamma \simeq 8 \times 10^{-5}$. Then using $\Omega_m \sim \Omega_c \sim 10^4\Omega_{\gamma}$, simple substitution gives that $m \sim 10^6$ separates fields that begin oscillations during the radiation and matter dominated eras.

We can estimate the contribution to the critical density today coming from ALPs  by assuming an instantaneous transition from $\Lambda$ to DM behaviour and redshifting the initial constant energy density from $a(t_{osc})$ to $a(t_0)=1$ as if it were ordinary CDM.  We have that:
\begin{align}
a_{osc} &= \left ( \frac{t_{eq}}{t_0} \right)^{1/6} \left( \frac{1}{m t_0} \right)^{1/2} ;\qquad m \gtrsim 10^6 \nonumber \\
a_{osc}&= \left ( \frac{1}{m t_0} \right)^{2/3}; \qquad m \lesssim 10^6 \nonumber \\
\label{eqn:aosc}
\end{align}
which leads to:
\begin{align}
\Omega_a &= \left( \frac{t_{eq}}{t_0} \right)^{1/2} \left( \frac{1}{t_0} \right)^{3/2} \frac{m^{1/2}}{6}\phi_0(t_i)^2;  \qquad m \gtrsim 10^6 \label{eqn:omegarad} \\
\Omega_a &= \frac{1}{6} \left( \frac{1}{t_0} \right)^2 \phi_0(t_i)^2; \qquad m \lesssim 10^6 \label{eqn:omegamat}
\end{align}
These expressions can be easily inverted to find an expression for the initial condition $\phi_0(t_i,\Omega_a)$. In our code we supplement these with an iteratively improved constant to take into account the ALP effects on $t_0$,  and $t_{eq}$.

There are two remaining quantities to be determined, if we take the initial scale factor, $a_i$ as an input parameter: $t_i$ and $t_0$. We begin in the radiation dominated era, so that:
\begin{equation}
t_i = \left( \frac{t_{\mathbf{eq}}}{t_0} \right)^{-1/3}a_i^2 t_0
\label{eqn:intitialt} 
\end{equation}
We know $t_0$ for a matter dominated universe: $t_0=2/3$. If we assume for now that ALP effects on the age of the universe away from what we expect from a pure matter or $\Lambda$ dominated universe are small (we will address this in more detail in Section~\ref{Background_Results}) then we can include the effect of $\Lambda$ on what initial conditions we will need by integrating Eqn.~\ref{eqn:hubble}. The calculation can be found in the standard textbooks \cite{book:dodelson, book:peacock}:
\begin{equation}
t_0 = \frac{2}{3} \Omega_{\Lambda}^{-1/2} \sinh ^{-1} \left ( \left ( \frac{\Omega_{\Lambda}}{\Omega_m} \right )^{1/2} \right )
\end{equation}
where $\Omega_m = \Omega_c + \Omega_a$, which is a monotonically increasing function of $\Omega_{\Lambda}$- a cosmological constant makes the universe older. 

The difference in our approach here for computing $\Omega_a$ from that in previous works that have been mainly concerned with the QCD axion, e.g. \cite{turner1986, fox2004}, is that we assume no temperature variation to the axion mass, or rather we assume that it has reached its zero temperature value quickly, and crucially before oscillations of the field begin, which since the fields are so light will be at a low temperature any how. We consider this a reasonable simplification because we do not in general know the temperature dependence of the mass for a string axion since we do not know what instantons will make the dominant contribution to the potential.

If we wanted to restrict our analysis to true axions, rather than the more general case of ultra-light scalar field ALPs, there is an important region of axion parameter space, known as the ``anthropic boundary'', which is instructive to locate. True axions are periodic in $\theta = \phi/f_a$ and so the initial misalignment angle has a ``maximum'' at $\theta_i = \pi$. In addition, for axions that begin oscillations in the radiation dominated era, $\Omega_a$ depends on the axion mass (Eqn.~\ref{eqn:omegarad}). The result is that for masses $m \lesssim 10^{12}$ it is impossible, without taking account of anharmonic terms in the potential and tuning the initial misalignment arbitrarily close to $\pi$ \cite{turner1986}, for axions to produce $\Omega_a >1$ and overclose the universe. This somewhat alleviates fine tuning problems for these light, high-$f_a$ ALPs that lead, in the usual case, to one having to tune $\theta(t_i)$ arbitrarily close to zero to prevent overclosure of the universe and other cosmological problems~\cite{dine1983, visinelli2009}. Therefore with ultra-light ALPs the fine tuning arguments of Mack and Steinhardt \cite{mack2009a, mack2009b} lose some of their power.

The flip side to this is that without tuning the initial misalignment arbitrarily close to $\pi$ and including anharmonic effects in the potential it is impossible for axions with masses much below the anthropic boundary to constitute an order one fraction of the dark matter. For example, axions that begin oscillating in the matter dominated era, such that $\Omega_a$ is independent of the mass (Eqn.~\ref{eqn:omegamat}), the maximum possible contribution to the energy budget is $\Omega_a \sim 4\times 10^{-4}$. This observation seems to cause problems for the arguments in \cite{tegmark2006} that any axion should contribute an order one fraction of the dark matter. 

These conflicting observations on fine tuning for axions below the anthropic boundary serve as further motivation for the discussion and subsequent decision given in Section~\ref{intro} to generalise to ultra-light scalar fields with a quadratic potential that do not respect the shift symmetry of axions when considering such low masses as are relevant for the FCDM scenario.

\section{Evolution of the Pertrubations}
\label{perturbations}

\subsection{Initial Conditions}

To solve for the evolution of the density perturbations and compute the resulting matter power spectrum we need to find the appropriate initial conditions for the perturbations in the various fluid components. 

The initial fluid perturbations can be separated into adiabatic and isocurvature components. Whilst axions are a source of isocurvature perturbations, and these play an important role in constraining axion models by their effect on the CMB (see, for example, \cite{visinelli2009, fox2004}), we will not consider them here, since they do not have a considerable bearing on the matter power spectrum.

Adiabatic perturbations are laid down in a scale invariant way after inflation, and occur in all fluid components from their coupling to the metric perturbation, $h$. Working to lowest order in $k\tau$ the coupled equations Eqns.~\ref{eqn:h},~\ref{eqn:deltag}, and \ref{eqn:thetag} can be solved analytically and the dominant late time growing mode solution is~\cite{bertschinger1995}:
\begin{align}
h &= C(k)(k\tau)^2 \\
\delta_{\gamma} &= -\frac{2}{3}C(k)(k\tau)^2 \\
\theta_{\gamma} &= -\frac{1}{18}C(k)(k^4\tau^3)
\end{align}
where $C(k)$ is fixed by the primordial power spectrum.  The $k$ dependence be fixed by assuming scale invariance, which requires $\delta \sim k^{1/2}$ and so $C(k)=Ck^{-3/2}$.

To find the initial condition on $\phi_1$ we use the condition of zero entropy relating adiabatic perturbations in two fluids $a$ and $b$:
\begin{align}
S_{ab} &= \frac{\delta_a}{1+w_a} - \frac{\delta_b}{1+w_b} \\
S_{ab} &= \dot{S}_{ab} = 0
\end{align}
From this one finds that the initial values of $\phi_1$ and $\dot{\phi}_1$ both depend on $\phi_{0,t}(0)$, the initial value of the derivative of $\phi_0$ with respect to physical time~\cite{perrotta1999}. For ultra-light scalar fields that are the PGBs of a spontaneously broken symmetry this derivative is zero, since the field is frozen at the initial misalignment  by Hubble drag, and there is no initial velocity. Therefore the initial conditions are simply:
\begin{align}
\phi_1 &= 0 \\
\dot{\phi}_1 &= 0
\end{align}

The initial conditions derived in~\cite{bertschinger1995} are for a numerical integration beginning in the radiation dominated era, where we also begin our simulations. We can integrate the background expansion, Eqn.~\ref{eqn:friedmann}, and find:
\begin{equation}
\tau_i = \frac{1}{\sqrt{\Omega_{\gamma}}}a_i
\end{equation}
which is the final input required to fix the initial conditions completely, up to the constant $C$ that sets the size of the initial perturbations, which we need not specify since we will only be considering ratios of power spectra.

\subsection{Suppression of Structure Formation}

Once the scalar field is deep into its oscillatory phase and the background evolution is well described by pure matter or radiation domination it is possible to solve the equations of motion with a WKB approximation. This gives:
\begin{align}
\phi_0(\tau) &= A_0 \left( \frac{a(\tau_{osc})}{a(\tau)} \right)^{3/2} \cos\left[  m \int_{\tau_{\mathbf{osc}}}^{\tau}\mathrm{d}\tau a(\tau)\right]  \\
\phi_1 (\tau) &= A_1 \left( \frac{a(\tau_{osc})}{a(\tau)}\right)^{3/2} \cos\left[ \int_{\tau_{\mathbf{osc}}}^{\tau} \mathrm{d}\tau (m^2 a(\tau)^2 + k^2)^{1/2}\right]
\end{align}
These solutions can then be averaged over their rapid oscillations and an expression for the sound speed~\cite{hu2000, hu1998b} is derived in two asymptotic regimes, as seen earlier in Eqn.~\ref{eqn:cssquared}. This momentum dependent sound speed leads to the emergence of a new scale in the scalar field evolution: $k_R = ma$. For $k<k_R$ the sound speed in the density perturbations is small and the scalar field behaves as ordinary dust CDM. For $k>k_R$ the sound speed in the density perturbations goes relativistic and the particles free-stream. This similarly defines a time $\tau_R$ after which a given mode ceases to behave relativistically. 

A mode of wavenumber $k$ crosses the horizon when $k\approx Ha$, or equivalently $k\tau \approx 1$. For a given $k$ this defines a time of horizon crossing $\tau_c$. Whether the free-streaming scale leads to a suppression of structure, and corresponding step-like feature in the matter power spectrum depends on the ordering of the times $\tau_c$ and $\tau_R$. If a given mode enters the horizon once it has already become non-relativistic, $\tau_R<\tau_c$, then the density perturbations in that mode will behave as ordinary CDM and there will be no suppression of structure relative to the standard model. However if a mode is relativistic when it enters the horizon, so that $\tau_c<\tau_R$ then structure will be suppressed on that scale. We estimate the scale $k_m$ at which suppression of structure formation begins for a field of a given mass, $m$, at a given redshift, $z$.

Smaller and smaller $k$ values are entering the horizon at all times. The mode that entered the horizon at matter-radiation equality corresponds to $k_c(z_{eq})\approx 0.03 h\unit{Mpc}^{-1}$. As the scale factor increases so does the boundary value for relativistic modes, $k_R$. We require $k_R<k_c$, for suppression of structure at a given redshift $z$. The large scales that we are interested in for the FCDM scenario suppress structure formation in modes that entered during matter domination. During matter domination $a~\tau^2$, which leads to a prediction for the mass dependence of the scale $k_m$:
\begin{equation}
k_m(m) \sim m^{1/3}
\label{eqn:km}
\end{equation}
For heavier scalar fields that suppress structure formation in modes that entered during the radiation dominated era we have:
\begin{equation}
 k_m \sim m^{1/2}
 \label{eqn:km2}
 \end{equation}
so that masses in the range $10^4\lesssim m \lesssim 10^6$ separate the regions. However we also know that only masses $m \gtrsim 10^6$ were oscillating in the radiation era, and therefore it is only for these masses that the derivation of $k_R$ holds, so that in addition we expect some numerical corrections and $z$ dependence to be introduced into the expressions above by the transitionary dynamics of the background expansion between matter and radiation domination, which will effect the expansion rate used to derive $k_m$, and due to the background scalar field transition between DM and $\Lambda$ behaviour. This will be most severe for fields that are still undergoing their transition at the redshift of observation, $z_{obs}$.

How much suppression of structure do we expect relative to ordinary CDM? The matter power spectrum is given by $P(k)=\delta_m^2$, where $\delta_m$ is the total overdensity in matter: $\delta_m=(\delta\rho_c+\delta\rho_a)/\rho_m$. For our purposes it will be useful to normalise the power spectrum to one on the largest scales. After matter radiation equality, density perturbations in ordinary CDM grow like $\delta\sim a$. It is a well known result~\cite{bond1980} that if a fraction $f(z)$ of the matter is unable to cluster then perturbations grow as $\delta\sim a^q$, where $q=1/4(-1+\sqrt{25-24f(z)})$. The deviation from $q=1$ at a given redshift will therefore start at the scale $k_m$ and saturate at the Jeans scale $k_J=a\sqrt{Hm}$: this is why we expect to see the emergence of ``steps'' in the power spectrum relative to ordinary CDM~\cite{axiverse2009, amendola2005}.

We can estimate the suppression of power, $S$, in a step using a parameterisation found in~\cite{amendola2005} found by taking the ratio of the two different growth rates of the density perturbations:
\begin{equation}
S(a)=\left( \frac{a_s}{a} \right)^{2(1-q)}
\label{eqn:sguess}
\end{equation}
where $a_s = max(a_{osc},a_{eq})$. The size of a step is then given by $1-S$.

An important difference between this work and the work in~\cite{amendola2005} is that we make no approximations for the evolution of the scalar field when actually computing $P(k)$.

\subsection{The Scales Involved}

Here we summarise the previous sections by restating the important scales to consider when thinking about the effects of ultra-light scalar fields on structure formation.
\begin{itemize}
\item A scalar field receives an initial value after symmetry breaking and at early times it remains frozen at this value by the Hubble drag. A frozen scalar field behaves as a cosmological constant; once it begins oscillating it will behave as matter. A field begins oscillating when:
\begin{equation*}
H(t) < m
\end{equation*}
\item Do oscillations begin in the radiation or matter dominated era? We will mostly be interested in ultra-light fields that begin oscillations in the matter dominated era:
\begin{equation*}
m \lesssim 10^{-27}\unit{eV}
\end{equation*}
\item The energy density today in such an ultra-light field depends on its initial value as:
\begin{equation*}
\Omega_a = \frac{1}{6} \left( \frac{1}{t_0} \right)^2 \phi_0(t_i)^2
\end{equation*}
\item Perturbations in the scalar field have a scale dependent sound speed, so we can ask: are the perturbations on a given scale at a given time relativistic? The scale $k_R=ma(t)$ separates the two regimes. On small scales:
\begin{equation*}
k>k_R
\end{equation*} 
the sound speed is relativistic and the particles \emph{free-stream}, suppressing structure formation.
\item Time dependence of the free-streaming scale and the finite size of the horizon mean that suppression of structure formation will accumulate on scales larger than the free-streaming scale. For the ultra-light fields under consideration, suppression of structure begins at a scale:
\begin{equation*}
k_m \sim \left( \frac{m}{10^{-33}\unit{eV}} \right)^{1/3}\left( \frac{100\unit{km}\unit{s}^{-1}}{c} \right) h\unit{Mpc}^{-1}
\end{equation*}
\item The steps in the power spectrum caused by this suppression of structure depend on the fraction, $f$, of matter in ultra-light fields. The amount of suppression can be estimated as:
\begin{equation*}
S(a)=\left( \frac{a_{osc}}{a} \right)^{2(1-1/4(-1+\sqrt{25-24f}))}
\end{equation*}
As one would expect, a larger $f$ gives rise to greater suppression of structure, as do lighter fields that free-stream on larger scales.
\end{itemize}

\section{Results in the Background}
\label{Background_Results}

\begin{figure}
\centering
\includegraphics[scale=0.4]{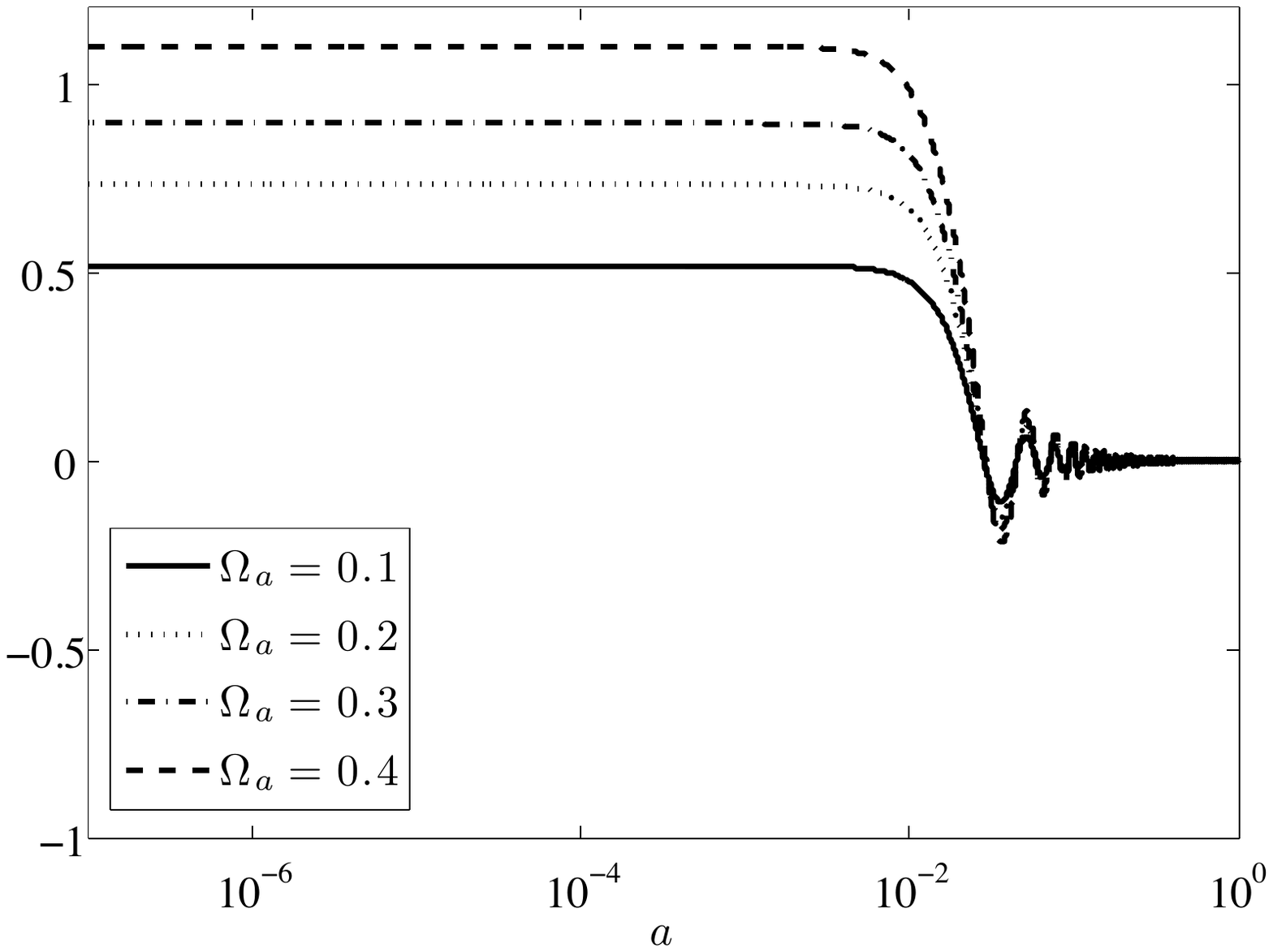}
\caption{$\phi_0$ versus $a$ for $m = 10^3$, various $\Omega_c$, $\Omega_{\Lambda} = 0$}
\label{fig:phi_o}
\end{figure}

Firstly we show a representative figure, Fig.~\ref{fig:phi_o}, for the evolution of the field $\phi_0$. This shows how the initial misalignment depends on $\Omega_a$. 

\begin{figure}
\centering
\includegraphics[scale=0.4]{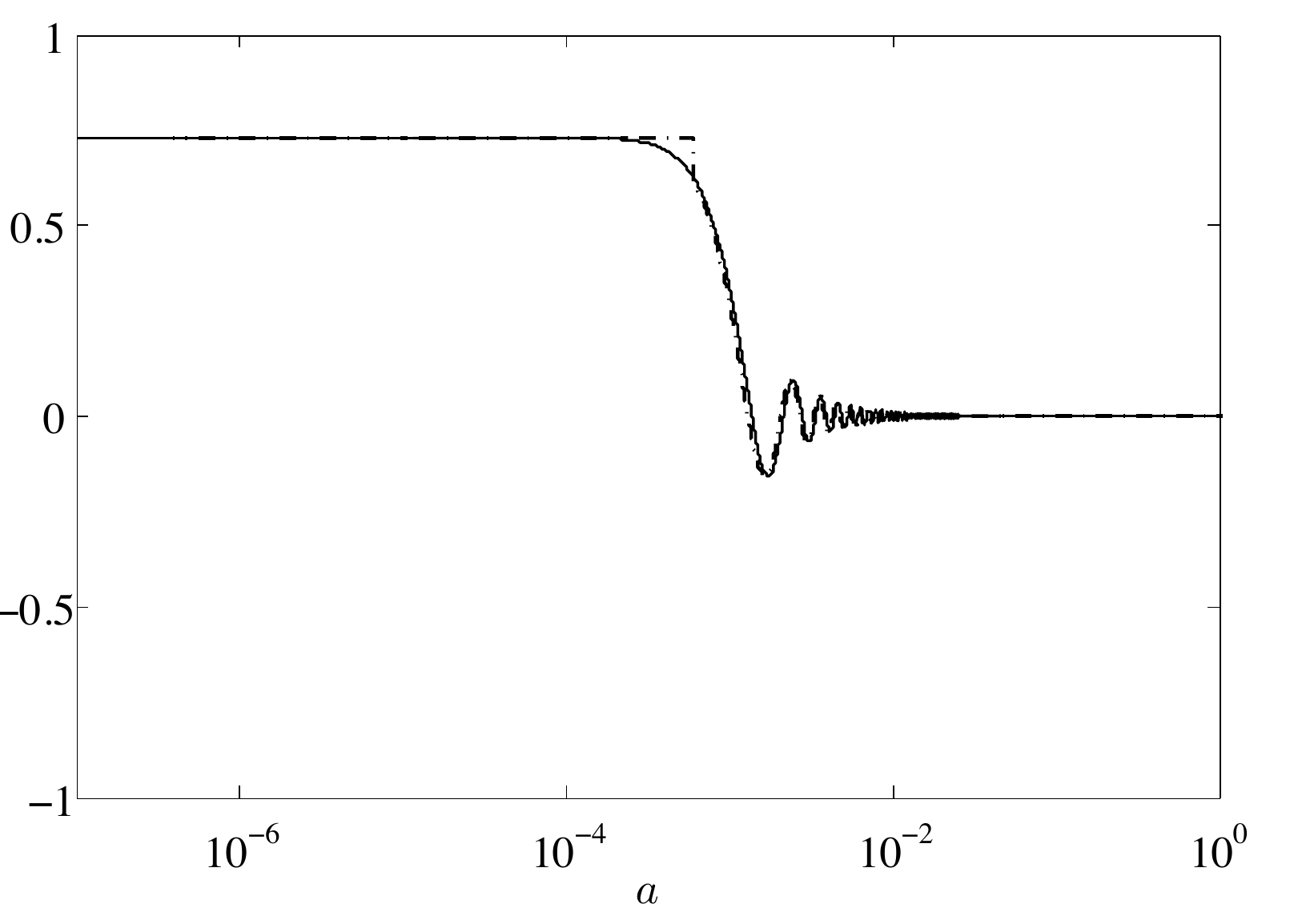}
\caption{A fit for $\phi_0$ versus $a$ for $m=10^5$, $\Omega_c = 0.8$, $\Omega_{\Lambda} = 0$. The solid line is the result of numerically integrating the equations of motion, whilst the dotted line is the analytic fit of Eqn.~\ref{eqn:phibessel}.}
\label{fig:phi_fit}
\end{figure}

In Fig.~\ref{fig:phi_fit} we show a representative fit to the numerical solutions using the analytic results obtained in Section~\ref{Background}. The fit is made by fixing the field to its initial value before $a_{osc}$, and then applying the analytic solution of Eqn.~\ref{eqn:phibessel} after $a_{osc}$. The analytic solution captures the decay envelope and transition well, and the estimate for $a_{osc}$ is also a good one.

\begin{figure}
\centering
\includegraphics[scale=0.4]{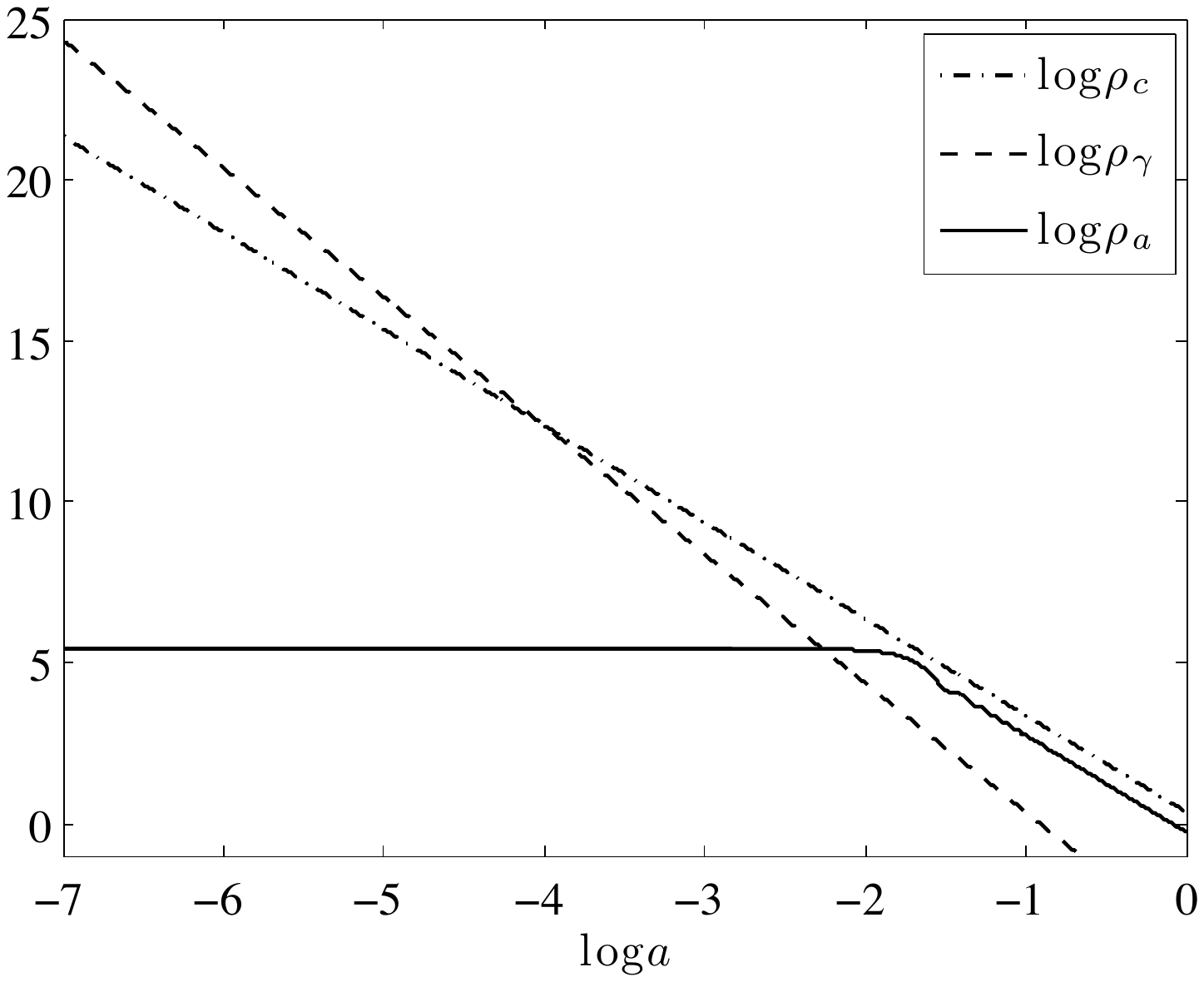}
\caption{$\log \rho$ versus $\log a$ for $m = 10^3$, $\Omega_c = 0.8$, $\Omega_{\Lambda} = 0$}
\label{fig:logrho}
\end{figure}

There are two important scales in the background evolution: the redshift of matter radiation equality, $z_{eq}$, and the redshift at which axion oscillations begin, $z_{osc}$. Both of these can be identified on a plot of $\log a$ versus $\log \rho$, Fig.~\ref{fig:logrho}, and agree well with the expected values, so that for example with $m = 10^3$ the oscillations begin in the matter dominated era.
Fig.~\ref{fig:logrho} again demonstrates the main features of the scalar field evolution, this time through the redshifting of the energy density: there is a period of constant energy density, and a period where the energy density redshifts with the matter. However there is also a significant region between $a\sim 10^{-2}$ and $a\sim 10^{-1}$ where the scalar field undergoes a transition in its behaviour. 

\begin{figure}
\centering
\includegraphics[scale=0.4]{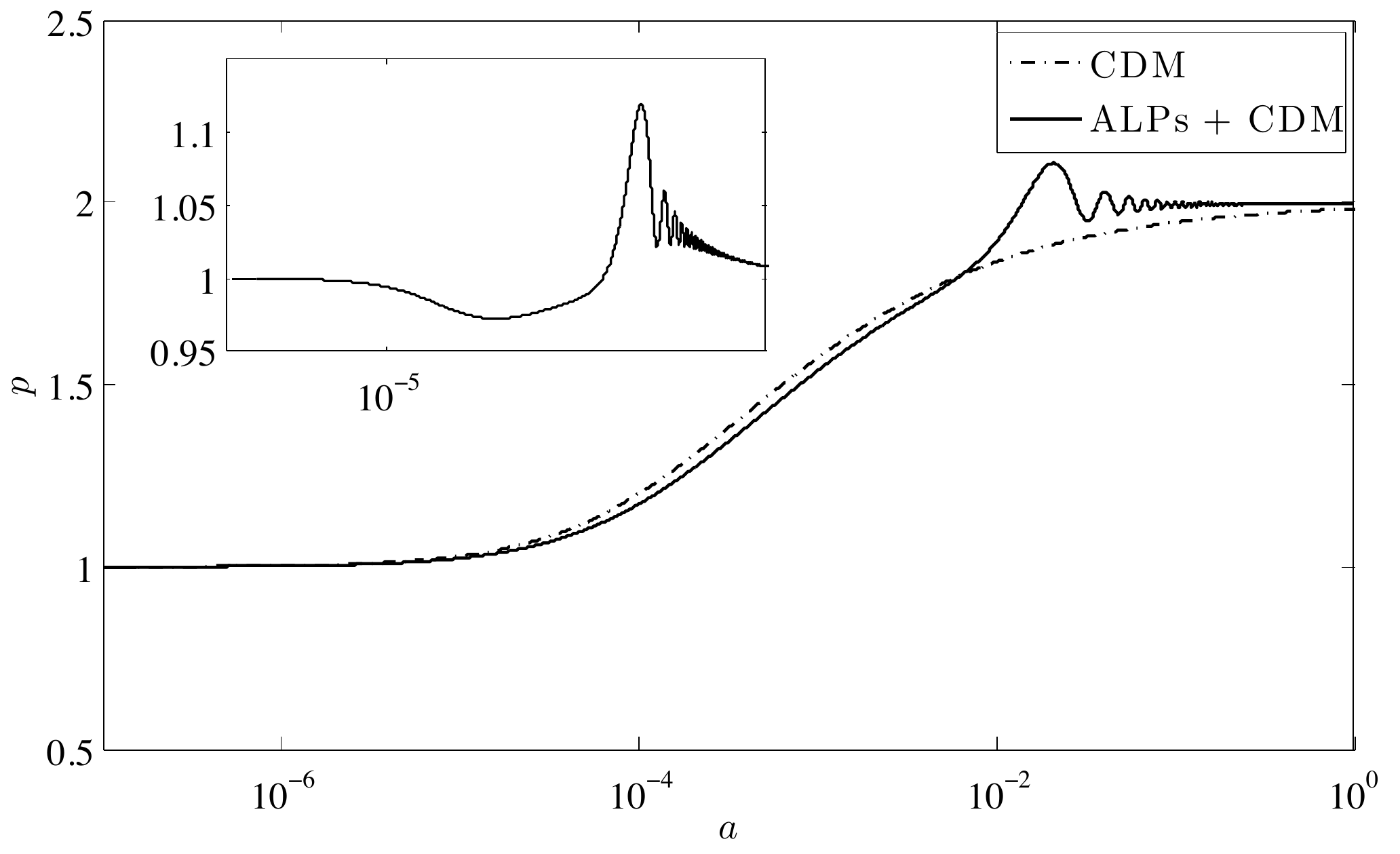}
\caption{$p$ versus $a$ for $m = 10^3$, $\Omega_c = 0.8$, $\Omega_{\Lambda} = 0$. Insert: ALPs $+$ CDM divided by standard CDM alone}
\label{fig:p}
\end{figure}

One way of looking at the length and significance of this transition is to see how it effects the expansion rate. In a pure matter, or pure radiation dominated era the scale factor evolves as $a\sim \tau^p$, with a slow, smooth transition between the two regimes (see Fig.~\ref{fig:p}). Therefore plotting $\mathcal{H}\tau$ versus scale factor will extract the time evolution of $p$. When the scalar field begins oscillation there is a more rapid time dependence introduced to $p$, however one can demonstrate that for the ultra-light scalar fields of interest here, where oscillations begin around $\tau \sim \mathcal{O}(1)$, that the product $\dot{p}\tau \ln \tau$ remains small compared to $p$ and therefore the expansion is still well described by plotting $\mathcal{H}\tau$ versus scale factor. A plot of the exponent $p$ during the transition in the scalar field behaviour estimated in this way is also shown in Fig.~\ref{fig:p}.  The relative change in the numerical value of $p$ is small during this transition. The transition here begins well into the matter epoch, and lasts for approximately one order of magnitude in scale factor growth, much shorter than the transition from pure radiation to pure matter expansion, which lasts over four orders of magnitude for the case at hand. Results for different masses of scalar field show that the length of the transition as one order of magnitude in scale factor is approximately mass independent. These two observations of a relatively short and small effect on the expansion rate due to the presence of an ultra-light scalar field lead to the prediction that such a dark matter component will have a correspondingly small effect on the age of the universe, and this intuition is indeed borne out in the numerical simulation.

When a non-zero cosmological constant is included the effect on the evolution of the background field is minimal. An accelerated expansion rate at late times is included simply by altering $\phi_0(t_i)$ through the effect on $t_0$ as discussed in Section~\ref{Background}: a larger $t_0$ will simply increase the required initial misalignment for a fixed fraction in ALPs; the accelerated expansion dilutes the ALP density. Furthermore, the mass range of ALPs that will contribute to the matter fraction, $\Omega_m$, today must be deep in their oscillatory regime. Those ALPs light enough that we would expect oscillations to begin in a $\Lambda$ dominated era cannot fulfil this requirement and are overdamped by the increased expansion rate, as such we will not be interested in them since they will not be contributing to the matter power spectrum and will simply be adding on to increase the effective value of $\Lambda$: the lightest fields we will consider are of mass $m\gtrsim 10^2$. We conclude that the overall effect of $\Lambda$ is exactly as for standard $\Lambda$CDM and is unaltered by the presence of an ALP component in the dark matter.

\section{The Matter Power Spectrum}
\label{Power_Spectrum}

In this section we return to using physical, dimensionful variables.

In this work, in order to have the smallest and simplest set of fits to make to our results, which will be most useful for comparison with observations, we will simply fit for $k_m$ at the centre of any step in the power spectrum, and take the saturation at $k_J$ as given. We will be interested in the function:
\begin{equation}
T^2(k)=\frac{P(k)_{\mbox{\small{ALPs + CDM}}}}{P(k)_{\mbox{\small{CDM}}}}
\end{equation}
from which we can define the step size:
\begin{equation}
S=T^2(k_{max }\gg k_m)
\end{equation}
where $k_{max}$ is the smallest scale of interest in the numerical or observational situation at hand. The dependence on $m$, $\Omega_a$, and $z$ is implicit. Later we will consider the effect of non-zero $\Omega_{\Lambda}$.

\begin{figure}
\centering
\includegraphics[scale=0.4]{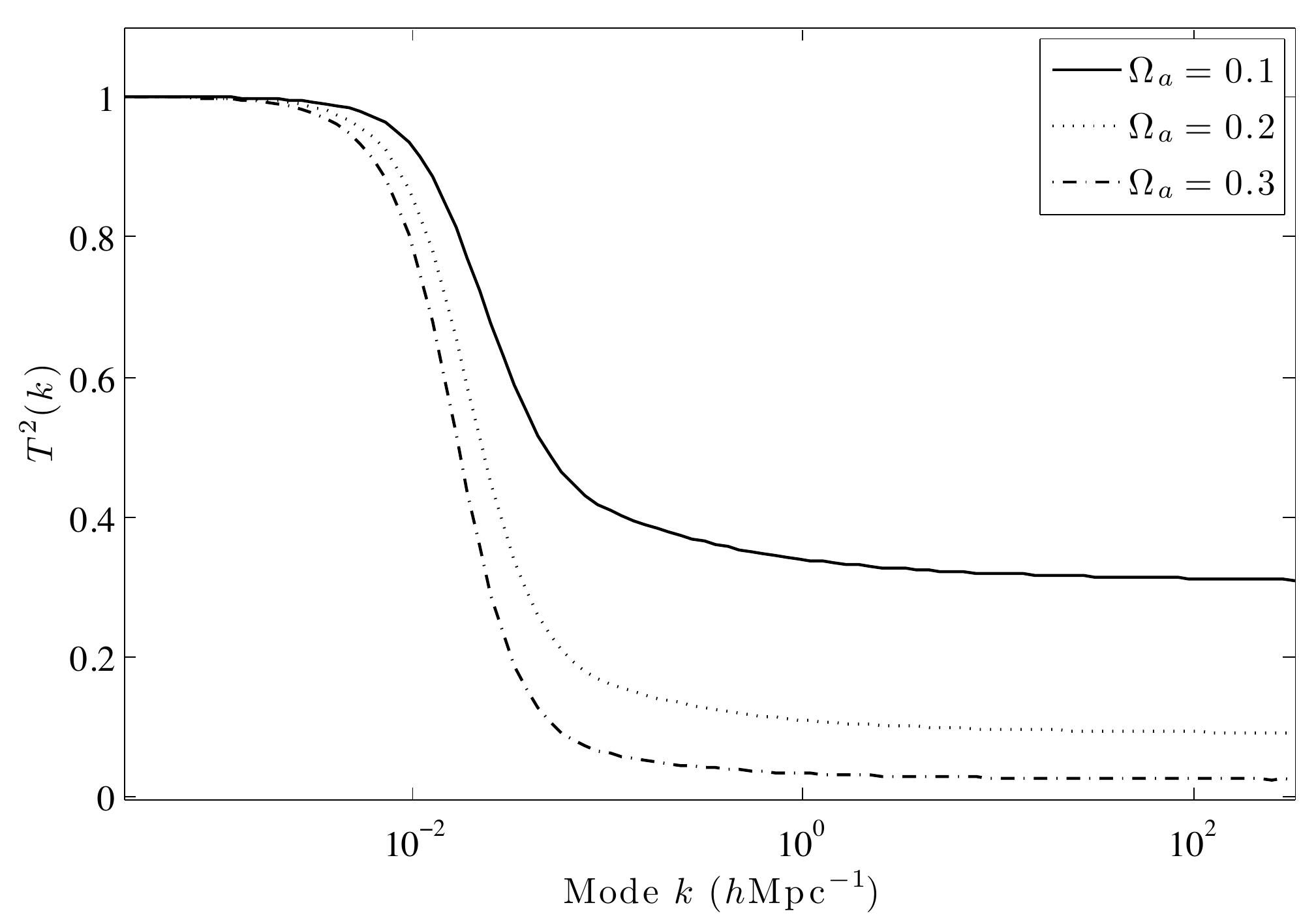}
\caption{$T^2(k)$ for $m = 10^{-29}\unit{eV}$, $\Omega_{\Lambda} = 0$}
\label{fig:steps_omega}
\end{figure}

In Fig.~\ref{fig:steps_omega} we show $T^2(k)$ for $m=10^{-29}$eV for various values of $\Omega_a$ at $z=0$. To fit for $S$ we use the parameterisation of Eqn.~\ref{eqn:sguess} for $T^2(k)$ taken at the largest $k=3\times 10^2$ $h$Mpc$^{-1}$ in our numerical results, modified with the addition of two exponents, $\beta_1$ and $\beta_2$:
\begin{equation}
S(z)=\left( \frac{(1+z)^{1+\beta_1}}{(1+z_{osc})^{1+\beta_2}} \right)^{2(1-q)}
\label{eqn:sfit}
\end{equation}
This fit applies when $z_{osc}<z_{eq}$ and when $z<z_J$, where $z_J$ is the redshift at which $k_J(z)=3\times 10^2$ $h$Mpc$^{-1}$. The factor of $q$ contains the dependence on $f(z)$, and $z_{osc}$ contains the dependence on the mass. The parameters $\beta_1$ and $\beta_2$ are chosen by trial and error, and take into consideration the approximations in the free-streaming scale and oscillation scale used to derive the fit. For those fields still far from matter-like behaviour at a given $z$, we expect a poorer fit. We also expect a poorer fit for those fields where $z_{osc}\sim z_{eq}$.

\begin{figure}
\centering
\includegraphics[scale=0.4]{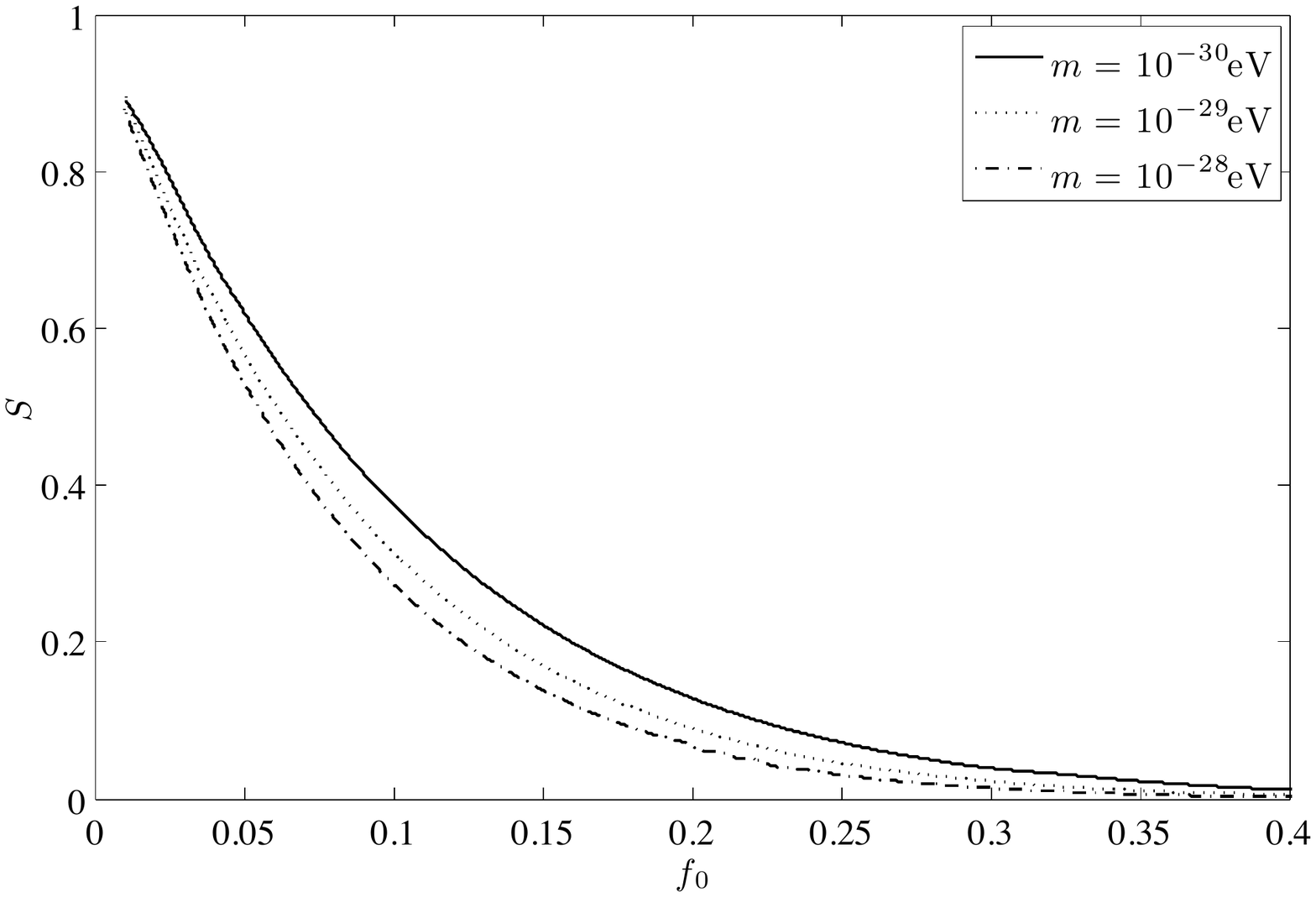}
\caption{$S$ versus $f(z=0) = f_0$ for $m = 10^{-30}$eV, $10^{-29}$eV, $10^{-28}$eV, $\Omega_{\Lambda} = 0$.}
\label{fig:stepsize_m}
\end{figure}

\begin{figure}
\centering
\includegraphics[scale=0.4]{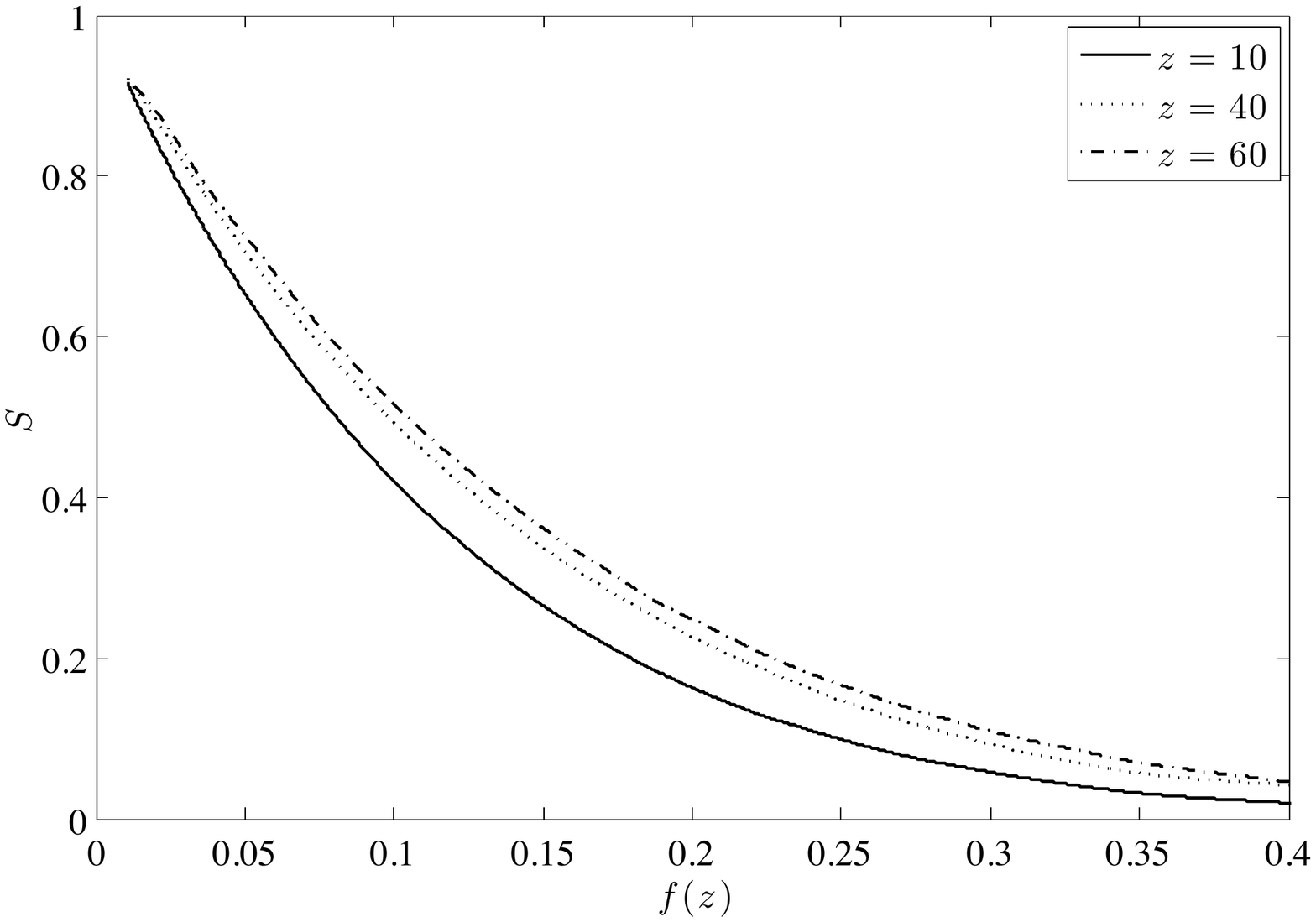}
\caption{$S$ versus $f(z)$ at $z=10$, $40$, $60$ for $m = 10^{-29}$eV, $\Omega_{\Lambda} = 0$.}
\label{fig:stepsize_z}
\end{figure}

In Fig.~\ref{fig:stepsize_m} we show $S(f)$ at $z=0$ for three masses of scalar field. The fits are used to determine $\beta_2$, and a reasonable match to within a few percent is obtained using the value $\beta_2=0.6$. Next, Fig.~\ref{fig:stepsize_z} shows $S(f(z))$ for $m=10^{-29}\unit{eV}$ with three values of $z$. The fits are again reasonable for a value of $\beta_1=0$, across two orders of magnitude in mass. Particularly, for $m=10^{-29}\unit{eV}$ the fit is good up to $z\sim 100$, but breaks down at $z\sim 200$, which is what we expect since at such high redshift the field is still very early in its transitionary regime and has yet to complete a full oscillation.

The final quantity we fit for is $k_m$, which we define by:
\begin{equation}
T^2(k_m)=\frac{1+S}{2}
\end{equation}
The fit should be well described by Eqn.~\ref{eqn:km} and only depend weakly on $z$ and $\Omega_a$ outside of the transition regime in the scalar field behaviour. For sufficiently small $z$ such that the field in question has completed its transition we use the fit:
\begin{equation}
k_m=A  f_0^{\alpha_1}(1+z)^{\alpha_2}m^{1/3}
\label{eqn:kmfit}
\end{equation}
where $f_0$ is the fraction in ALPs at $z=0$, $A$ is a constant of proportionality, and $\alpha_1$, $\alpha_2$ are exponents to be fit for. 

\begin{figure}
\centering
\includegraphics[scale=0.4]{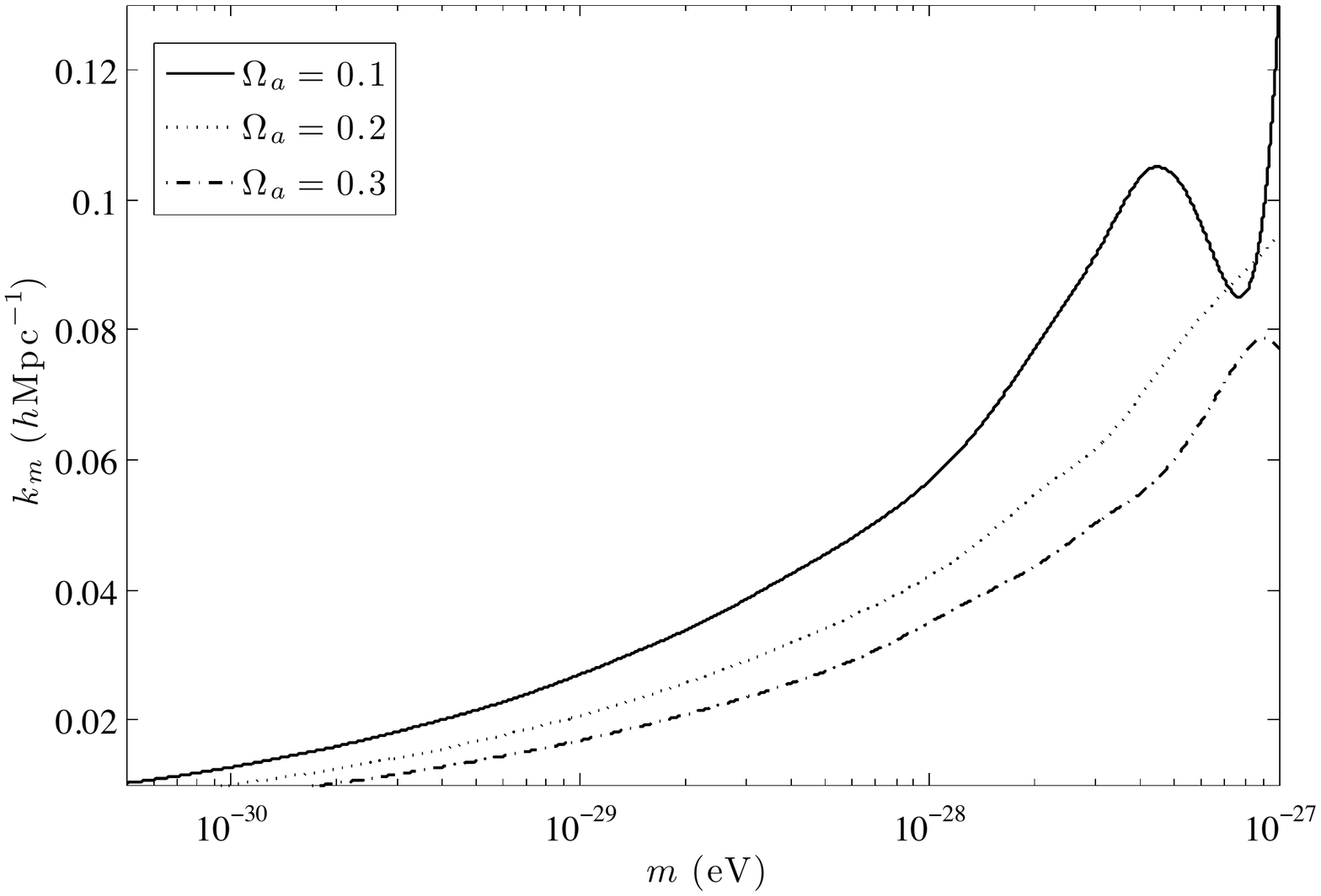}
\caption{$k_m$ versus $m$ at $z=0$ for $\Omega_a=0.1$, $\Omega_a=0.2$, $\Omega_a=0.3$, $\Omega_{\Lambda} = 0$.}
\label{fig:kmfit_o}
\end{figure}

Fig.~\ref{fig:kmfit_o} shows the shape of $k_m$ as a function of $m$ at $z=0$ for three values of $\Omega_a$. At $z=0$ all the masses of scalar field under consideration have, for the best part, undergone their full transition to matter like behaviour, and the shape is well fitted by Eqn.~\ref{eqn:kmfit} with $A=1.25$, $\alpha_1=-0.5$. With these values essentially normalising each curve, the shape follows $m^{1/3}$ well, until $m \gtrsim 10^{-28}$eV. The discussion surrounding Eqns.~\ref{eqn:km} and \ref{eqn:km2} tells us that these masses suppress structure formation in modes that entered when the background expansion rate was transitioning from pure radiation to pure matter dominated behaviour. In addition, at the time when the associated modes entered the horizon the masses in question were still undergoing their own transition from cosmological constant to dark matter behaviour, which will not only effect the background expansion rate but also the growth of scalar field overdensities so that neither Eqn.~\ref{eqn:km} nor Eqn.~\ref{eqn:km2} will hold exactly for the mass dependence of $k_m$. This departure from the naive picture underlines the importance of full numerical simulation to determine the effect of ultra-light scalar fields on the matter power spectrum.

\begin{figure}
\centering
\includegraphics[scale=0.4]{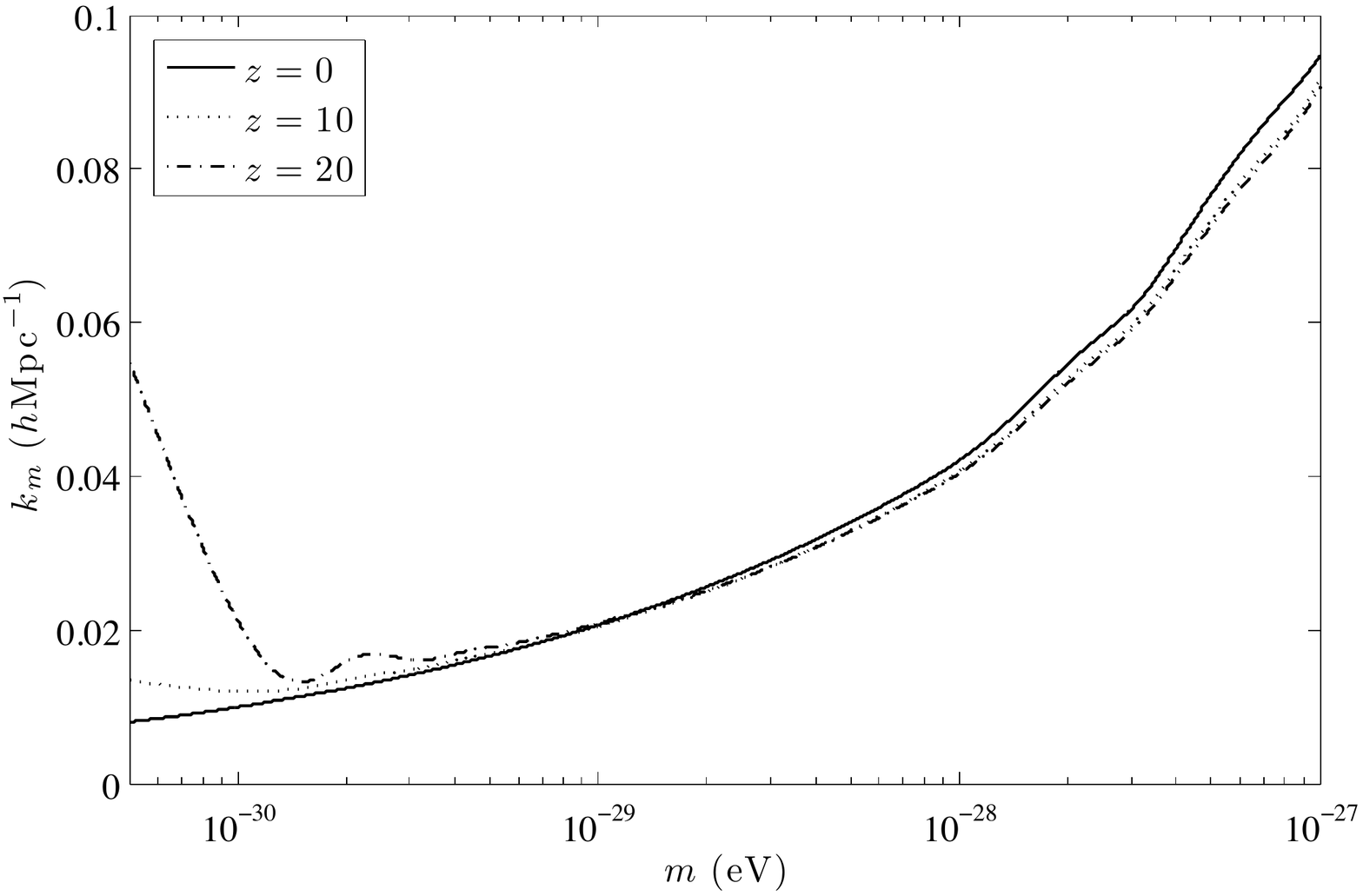}
\caption{$k_m$ versus $m$ at $z=0$, $10$, $20$ for $\Omega_c=0.8$, $\Omega_{\Lambda} = 0$.}
\label{fig:kmfit_z}
\end{figure}

Fig.~\ref{fig:kmfit_z} shows the shape of $k_m$ as a function of $m$ at three different $z$ values. When a scalar field is transitioning in behaviour from cosmological constant to CDM, or indeed when it is still behaving as pure cosmological constant, then the sound speed of Eqn.~\ref{eqn:cssquared} no longer applies and the steps in the power spectrum cannot be simply described by Eqn.~\ref{eqn:km}. At high redshift the lower mass scalar fields cause the curve to peel away from the $m^{1/3}$ shape. The fit of Eqn.~\ref{eqn:kmfit} does not attempt to capture the transitionary behaviour, and we only note that for fields with a larger mass that have completed the transition to CDM there is no remaining $z$ dependence in the location of the step, i.e. $\alpha_2=0$.

\begin{figure}
\centering
\includegraphics[scale=0.4]{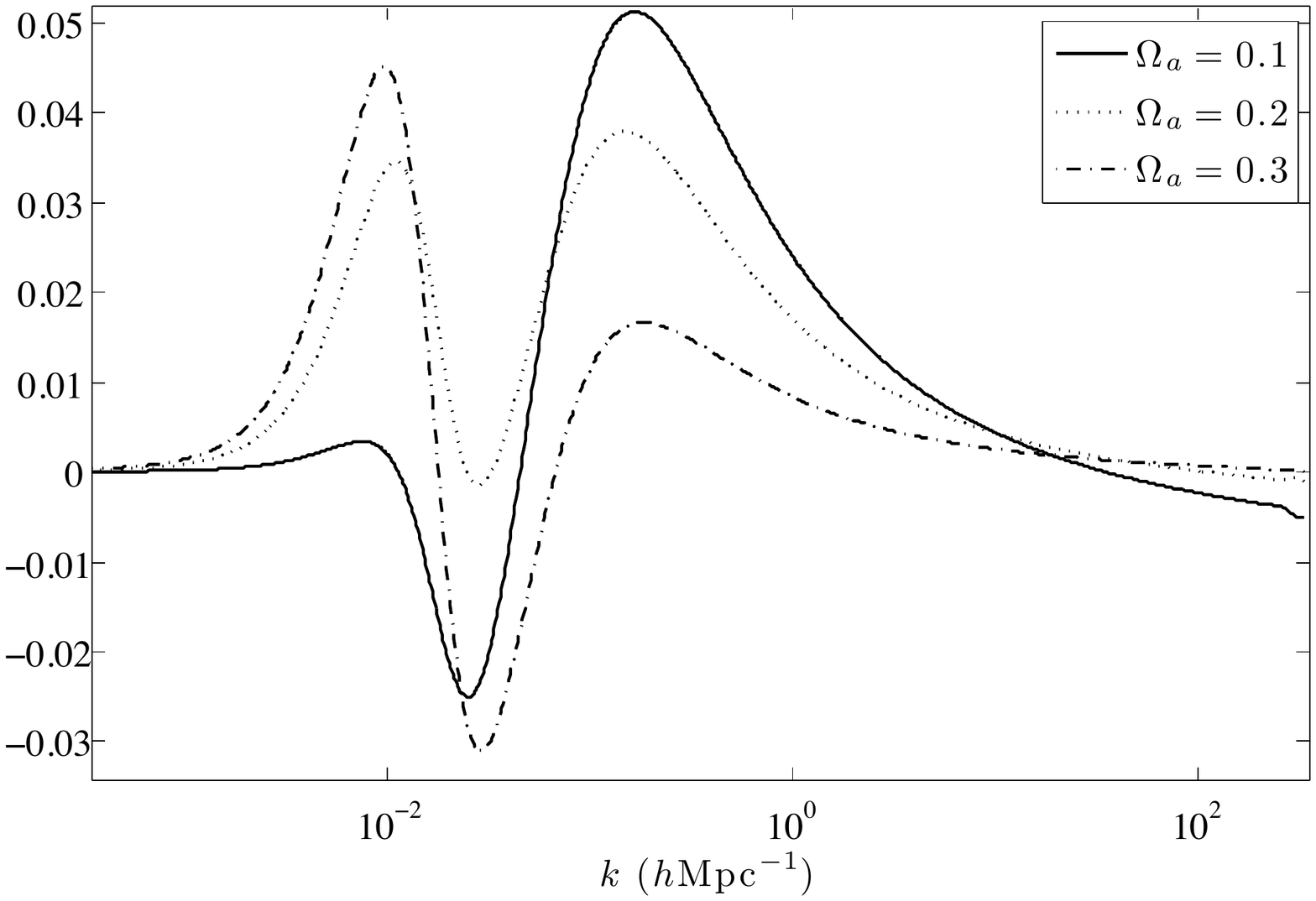}
\caption{Difference plot for the analytic fit of Eqn.~\ref{eqn:tfit} with $\gamma=2$ to the numerical results shown in Fig.~\ref{fig:steps_omega}}
\label{fig:stepfit}
\end{figure}

Given the step size, and the location of the centre of the step it is simple to fit the shape of $T^2(k)$:
\begin{equation}
T^2(k)=\frac{1+S(k/k_m)^{\gamma}}{1+(k/k_m)^{\gamma}}
\label{eqn:tfit}
\end{equation}
The difference plot for this fit against the steps shown in Fig.~\ref{fig:steps_omega} with $\gamma=2$ is shown in Fig.~\ref{fig:stepfit}, and is good within around $5\%$ of the total step size, $1-S$. The fit is worst around $k_m$, where we expect the most uncertainty in our fits, both from the discussion of the fit to Eqn.~\ref{eqn:kmfit}, and from our choice to fit for the centre of the step only, ignoring the difference between $k_m$ and $k_J$.

\begin{figure}
\centering
\includegraphics[scale=0.4]{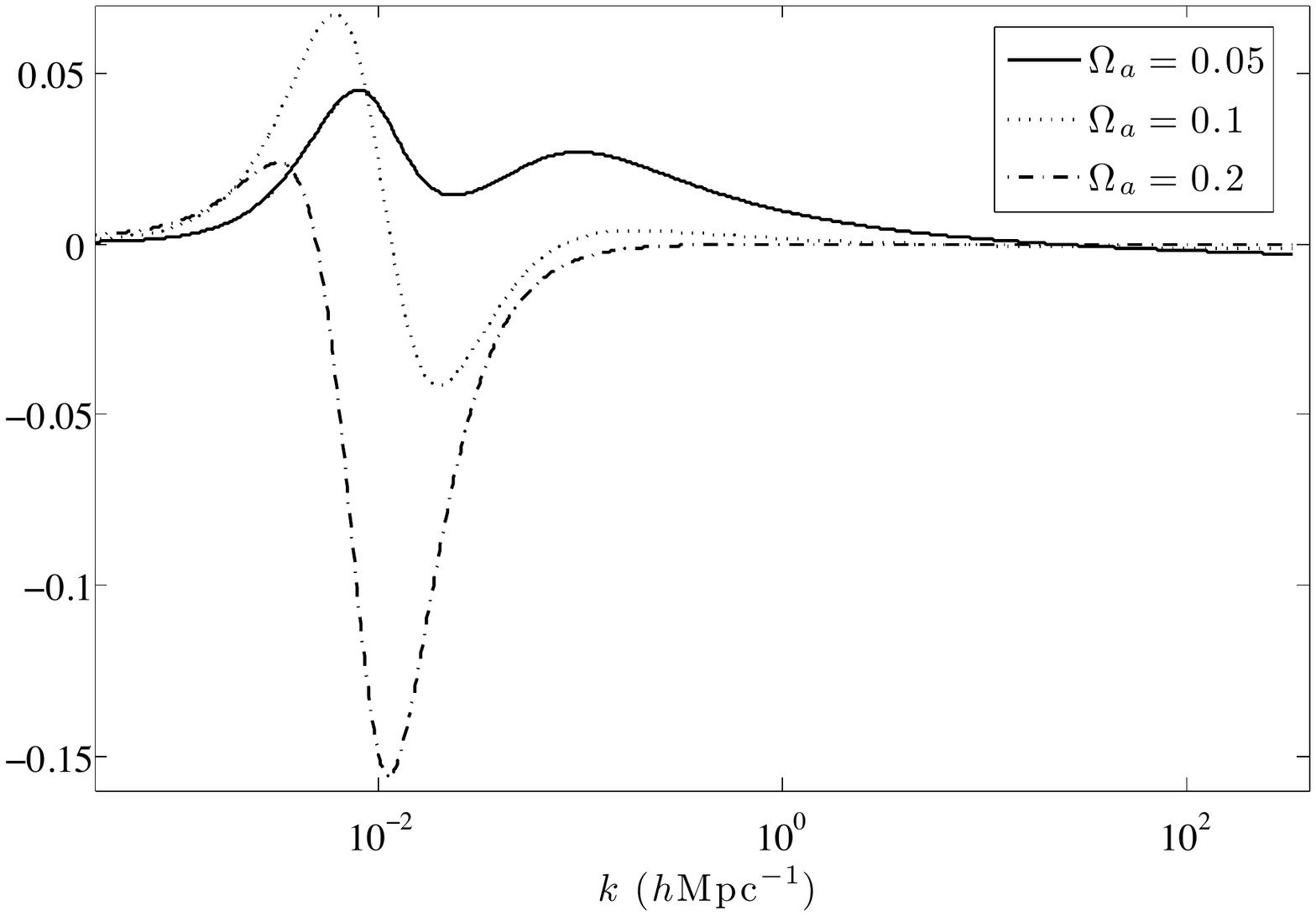}
\caption{Difference plot for the analytic fit of Eqn.~\ref{eqn:tfit} with $\gamma=2$, using Eqn.~\ref{eqn:kmfitL} with $\alpha_3=0.75$ to fit $k_m$, to numerically generated power spectra for $m=10^{-29}$eV and $\Omega_{\Lambda}=0.7$ at $z=0$.}
\label{fig:stepfitL}
\end{figure}

When a non-zero cosmological constant is included corresponding to $\Omega_{\Lambda}=0.7$, with no change to the fits, they remain good at large $k$ and so we conclude that $S$ has little or no dependence on $\Omega_{\Lambda}$. The fit is very poor for $k\sim k_m$ so we modify Eqn.~\ref{eqn:kmfit} to include one more exponent, $\alpha_3$:
\begin{equation}
k_m=A  f_0^{\alpha_1}(1+z)^{\alpha_2} (1-\Omega_{\Lambda})^{\alpha_3} m^{1/3}
\label{eqn:kmfitL}
\end{equation}
With $\alpha_3=0.4$ we return to fits good to around $5 - 15\%$ of the total step size, as shown in Fig.~\ref{fig:stepfitL}.

We have not considered power spectra for fields with $m>10^{-27}\unit{eV}$ that begin oscillating in the radiation era, and suppress structure formation in modes that entered during this era, since their behaviour will be much simpler: scalar fields with these masses will be much better fit by the approximations of~\cite{amendola2005}, and of our fits, in almost all observational situations. They will not effect the epoch of equality, and will only effect the expansion rate beyond $z_{eq}$, and, because they will be well transitioned to DM behaviour at all redshifts of interest then the scale $k_m$ will be extremely well fit by $m^{1/2}$.

\section{Discussion}
\label{Discussion}

In this paper we have explored the cosmological behaviour of ultra-light scalar fields and found that they can have a significant effect on the growth of structure at late times. This is not surprising given that we are essentially studying the clustering of very light axions, i.e. axions whose mass scale is close to the current cosmological
horizon.

Ultra-light axions have, until now, been difficult to motivate theoretically. Most of the focus of previous research has been on QCD or QCD-like axions for which there are well defined production mechanisms in the early Universe. There have been some studies of what has been dubbed Fuzzy Cold Dark Matter~\cite{hu2000}, albeit in a slightly different mass regime. But with the rise of the string axiverse~\cite{axiverse2009}, \cite{acharya2010} it makes sense to loosen our usual assumptions that tie the axion fractional energy, $\Omega_a$, to the mass scale, $m$. This means that we have been generous with what we deem allowable for the misalignment angle- to consider appreciable $\Omega_a$s we must allow large misalignment angles, pushing the dynamics of the axion field towards the anharmonic part of the axion potential. 

We have studied scalar fields whose evolution is such that the length scale for oscillations and clustering is greater than that set by the horizon size at radiation-matter equality. Hence, the transition from ``frozen'' (cosmological constant) to oscillatory behaviour occurs after radiation-matter equality. Two interesting new features arise. First of all, the energy density of the scalar field behaves as a dark energy at early times and then transits to dust like behaviour at late times- this transition happens after equality. Second, and as result of the previous point, the redshift of equality is shifted by a factor $(1-f)$. If $f$ is substantial, the affect can be appreciable. This transition between dark energy to dust like behaviour will affect the expansion rate and, unlike the standard axion picture, it will not be purely dust dominated. This effect can be of order $10\%$ for $\Omega_a\simeq 0.3$.

When we look at the evolution of perturbations and their effect on clustering we find that, as was expected, there is suppression of power on the smaller scales. The transition wave number, $k_m\sim m^{1/3}$, a mass dependence which is specific to these ultra-light fields, i.e. with a transition after radiation-matter equality. We have refined the dependence of the scale at which the suppression kicks in as well as the amount of the damping on small scales, giving us a better handle on how this effect depends on parameters such as $m$ and $\Omega_a$.

The next obvious step is to see if these various effects can be picked out in cosmological observables. These ultra-light fields affect the growth rate of perturbations and should be detectable by the standard selection of methods: the Integrated Sachs-Wolfe effect in the CMB, weak-lensing of the large scale structure and galaxy, cluster and Lyman-$\alpha$ surveys. An intriguing possibility is that these effects may contaminate and bias characteristic scales in large scale structure such as the Baryon Acoustic Oscillation. We will study these different effects in a future publication.




\section*{Acknowledgements}
\vspace{-10pt}
DM thanks Rhys Davies and John March-Russell for helpful discussions about string axions, Edward Macaulay for computing assistance, and Ren\'{e}e Hlozek for a careful reading of the manuscript and helpful discussions about the code. DM acknowledges support from an STFC Postgraduate Studentship and the EU Marie Curie Network UniverseNet (HPRN-CT-2006-035863). PGF acknowledges support from STFC, BIPAC and the Oxford Martin School.


\bibliographystyle{apsrev}
\bibliography{axions}

\begin{thebibliography}{36}
\expandafter\ifx\csname natexlab\endcsname\relax\def\natexlab#1{#1}\fi
\expandafter\ifx\csname bibnamefont\endcsname\relax
  \def\bibnamefont#1{#1}\fi
\expandafter\ifx\csname bibfnamefont\endcsname\relax
  \def\bibfnamefont#1{#1}\fi
\expandafter\ifx\csname citenamefont\endcsname\relax
  \def\citenamefont#1{#1}\fi
\expandafter\ifx\csname url\endcsname\relax
  \def\url#1{\texttt{#1}}\fi
\expandafter\ifx\csname urlprefix\endcsname\relax\def\urlprefix{URL }\fi
\providecommand{\bibinfo}[2]{#2}
\providecommand{\eprint}[2][]{\url{#2}}

\bibitem[{\citenamefont{Svrcek and Witten}(2006)}]{witten2006}
\bibinfo{author}{\bibfnamefont{P.}~\bibnamefont{Svrcek}} \bibnamefont{and}
  \bibinfo{author}{\bibfnamefont{E.}~\bibnamefont{Witten}}
  (\bibinfo{year}{2006}), \eprint{hep-th/0605206},
  \urlprefix\url{http://dx.doi.org/10.1088/1126-6708/2006/06/051}.

\bibitem[{\citenamefont{Peccei and Quinn}(1977)}]{pecceiquinn1977}
\bibinfo{author}{\bibfnamefont{R.~D.} \bibnamefont{Peccei}} \bibnamefont{and}
  \bibinfo{author}{\bibfnamefont{H.~R.} \bibnamefont{Quinn}},
  \bibinfo{journal}{Physical Review Letters} \textbf{\bibinfo{volume}{38}},
  \bibinfo{pages}{1440} (\bibinfo{year}{1977}),
  \urlprefix\url{http://dx.doi.org/10.1103/PhysRevLett.38.1440}.

\bibitem[{\citenamefont{Hooft}(1976{\natexlab{a}})}]{thooft1976a}
\bibinfo{author}{\bibfnamefont{G.~T.} \bibnamefont{Hooft}},
  \bibinfo{journal}{Physical Review Letters} \textbf{\bibinfo{volume}{37}},
  \bibinfo{pages}{8+} (\bibinfo{year}{1976}{\natexlab{a}}),
  \urlprefix\url{http://dx.doi.org/10.1103/PhysRevLett.37.8}.

\bibitem[{\citenamefont{Hooft}(1976{\natexlab{b}})}]{thooft1976b}
\bibinfo{author}{\bibfnamefont{G.~T.} \bibnamefont{Hooft}},
  \bibinfo{journal}{Physical Review D} \textbf{\bibinfo{volume}{14}},
  \bibinfo{pages}{3432} (\bibinfo{year}{1976}{\natexlab{b}}),
  \urlprefix\url{http://dx.doi.org/10.1103/PhysRevD.14.3432}.

\bibitem[{\citenamefont{Dine}(1981)}]{dine1981}
\bibinfo{author}{\bibfnamefont{M.}~\bibnamefont{Dine}},
  \bibinfo{journal}{Physics Letters B} \textbf{\bibinfo{volume}{104}},
  \bibinfo{pages}{199} (\bibinfo{year}{1981}), ISSN \bibinfo{issn}{03702693},
  \urlprefix\url{http://dx.doi.org/10.1016/0370-2693(81)90590-6}.

\bibitem[{\citenamefont{Preskill et~al.}(1983)\citenamefont{Preskill, Wise, and
  Wilczek}}]{preskill1983}
\bibinfo{author}{\bibfnamefont{J.}~\bibnamefont{Preskill}},
  \bibinfo{author}{\bibfnamefont{M.}~\bibnamefont{Wise}}, \bibnamefont{and}
  \bibinfo{author}{\bibfnamefont{F.}~\bibnamefont{Wilczek}},
  \bibinfo{journal}{Physics Letters B} \textbf{\bibinfo{volume}{120}},
  \bibinfo{pages}{127} (\bibinfo{year}{1983}), ISSN \bibinfo{issn}{03702693},
  \urlprefix\url{http://dx.doi.org/10.1016/0370-2693(83)90637-8}.

\bibitem[{\citenamefont{Steinhardt and Turner}(1983)}]{steinhardt1983}
\bibinfo{author}{\bibfnamefont{P.}~\bibnamefont{Steinhardt}} \bibnamefont{and}
  \bibinfo{author}{\bibfnamefont{M.}~\bibnamefont{Turner}},
  \bibinfo{journal}{Physics Letters B} \textbf{\bibinfo{volume}{129}},
  \bibinfo{pages}{51} (\bibinfo{year}{1983}), ISSN \bibinfo{issn}{03702693},
  \urlprefix\url{http://dx.doi.org/10.1016/0370-2693(83)90727-X}.

\bibitem[{\citenamefont{Turner et~al.}(1983)\citenamefont{Turner, Wilczek, and
  Zee}}]{turner1983}
\bibinfo{author}{\bibfnamefont{M.}~\bibnamefont{Turner}},
  \bibinfo{author}{\bibfnamefont{F.}~\bibnamefont{Wilczek}}, \bibnamefont{and}
  \bibinfo{author}{\bibfnamefont{A.}~\bibnamefont{Zee}},
  \bibinfo{journal}{Physics Letters B} \textbf{\bibinfo{volume}{125}},
  \bibinfo{pages}{35} (\bibinfo{year}{1983}), ISSN \bibinfo{issn}{03702693},
  \urlprefix\url{http://dx.doi.org/10.1016/0370-2693(83)91229-7}.

\bibitem[{\citenamefont{Abbott and Sikivie}(1983)}]{abbott1983}
\bibinfo{author}{\bibfnamefont{L.}~\bibnamefont{Abbott}} \bibnamefont{and}
  \bibinfo{author}{\bibfnamefont{P.}~\bibnamefont{Sikivie}},
  \bibinfo{journal}{Physics Letters B} \textbf{\bibinfo{volume}{120}},
  \bibinfo{pages}{133} (\bibinfo{year}{1983}), ISSN \bibinfo{issn}{03702693},
  \urlprefix\url{http://dx.doi.org/10.1016/0370-2693(83)90638-X}.

\bibitem[{\citenamefont{Dine and Fischler}(1983)}]{dine1983}
\bibinfo{author}{\bibfnamefont{M.}~\bibnamefont{Dine}} \bibnamefont{and}
  \bibinfo{author}{\bibfnamefont{W.}~\bibnamefont{Fischler}},
  \bibinfo{journal}{Physics Letters B} \textbf{\bibinfo{volume}{120}},
  \bibinfo{pages}{137} (\bibinfo{year}{1983}), ISSN \bibinfo{issn}{03702693},
  \urlprefix\url{http://dx.doi.org/10.1016/0370-2693(83)90639-1}.

\bibitem[{\citenamefont{Turner}(1986)}]{turner1986}
\bibinfo{author}{\bibfnamefont{M.~S.} \bibnamefont{Turner}},
  \bibinfo{journal}{Physical Review D} \textbf{\bibinfo{volume}{33}},
  \bibinfo{pages}{889} (\bibinfo{year}{1986}),
  \urlprefix\url{http://dx.doi.org/10.1103/PhysRevD.33.889}.

\bibitem[{\citenamefont{Visinelli and Gondolo}(2009)}]{visinelli2009}
\bibinfo{author}{\bibfnamefont{L.}~\bibnamefont{Visinelli}} \bibnamefont{and}
  \bibinfo{author}{\bibfnamefont{P.}~\bibnamefont{Gondolo}}
  (\bibinfo{year}{2009}), \eprint{0903.4377},
  \urlprefix\url{http://dx.doi.org/10.1103/PhysRevD.80.035024}.

\bibitem[{\citenamefont{Banks and Dine}(1996)}]{banks1996}
\bibinfo{author}{\bibfnamefont{T.}~\bibnamefont{Banks}} \bibnamefont{and}
  \bibinfo{author}{\bibfnamefont{M.}~\bibnamefont{Dine}}
  (\bibinfo{year}{1996}), \eprint{hep-th/9608197},
  \urlprefix\url{http://dx.doi.org/10.1016/S0550-3213(97)00413-6}.

\bibitem[{\citenamefont{Arvanitaki et~al.}(2009)\citenamefont{Arvanitaki,
  Dimopoulos, Dubovsky, Kaloper, and March-Russell}}]{axiverse2009}
\bibinfo{author}{\bibfnamefont{A.}~\bibnamefont{Arvanitaki}},
  \bibinfo{author}{\bibfnamefont{S.}~\bibnamefont{Dimopoulos}},
  \bibinfo{author}{\bibfnamefont{S.}~\bibnamefont{Dubovsky}},
  \bibinfo{author}{\bibfnamefont{N.}~\bibnamefont{Kaloper}}, \bibnamefont{and}
  \bibinfo{author}{\bibfnamefont{J.}~\bibnamefont{March-Russell}}
  (\bibinfo{year}{2009}), \eprint{0905.4720},
  \urlprefix\url{http://arxiv.org/abs/0905.4720}.

\bibitem[{\citenamefont{Linde}(1991)}]{linde1991}
\bibinfo{author}{\bibfnamefont{A.}~\bibnamefont{Linde}},
  \bibinfo{journal}{Physics Letters B} \textbf{\bibinfo{volume}{259}},
  \bibinfo{pages}{38} (\bibinfo{year}{1991}), ISSN \bibinfo{issn}{03702693},
  \urlprefix\url{http://dx.doi.org/10.1016/0370-2693(91)90130-I}.

\bibitem[{\citenamefont{Hertzberg et~al.}(2010)\citenamefont{Hertzberg,
  Tegmark, and Wilczek}}]{hertzberg2008}
\bibinfo{author}{\bibfnamefont{M.~P.} \bibnamefont{Hertzberg}},
  \bibinfo{author}{\bibfnamefont{M.}~\bibnamefont{Tegmark}}, \bibnamefont{and}
  \bibinfo{author}{\bibfnamefont{F.}~\bibnamefont{Wilczek}},
  \bibinfo{journal}{Physical Review D (Particles, Fields, Gravitation, and
  Cosmology)} \textbf{\bibinfo{volume}{78}}, \bibinfo{pages}{083507+}
  (\bibinfo{year}{2010}), \eprint{0807.1726},
  \urlprefix\url{http://dx.doi.org/10.1103/PhysRevD.78.083507}.

\bibitem[{\citenamefont{Sikivie}(2008)}]{sikivie2008}
\bibinfo{author}{\bibfnamefont{P.}~\bibnamefont{Sikivie}}, in
  \emph{\bibinfo{booktitle}{Axions}}, edited by
  \bibinfo{editor}{\bibfnamefont{M.}~\bibnamefont{Kuster}},
  \bibinfo{editor}{\bibfnamefont{G.}~\bibnamefont{Raffelt}}, \bibnamefont{and}
  \bibinfo{editor}{\bibfnamefont{B.}~\bibnamefont{Beltr\'{a}n}}
  (\bibinfo{publisher}{Springer Berlin Heidelberg}, \bibinfo{address}{Berlin,
  Heidelberg}, \bibinfo{year}{2008}), vol. \bibinfo{volume}{741} of
  \emph{\bibinfo{series}{Lecture Notes in Physics}},
  chap.~\bibinfo{chapter}{2}, pp. \bibinfo{pages}{19--50}, ISBN
  \bibinfo{isbn}{978-3-540-73517-5},
  \urlprefix\url{http://dx.doi.org/10.1007/978-3-540-73518-2\_2}.

\bibitem[{\citenamefont{Hu et~al.}(2000)\citenamefont{Hu, Barkana, and
  Gruzinov}}]{hu2000}
\bibinfo{author}{\bibfnamefont{W.}~\bibnamefont{Hu}},
  \bibinfo{author}{\bibfnamefont{R.}~\bibnamefont{Barkana}}, \bibnamefont{and}
  \bibinfo{author}{\bibfnamefont{A.}~\bibnamefont{Gruzinov}},
  \bibinfo{journal}{Physical Review Letters} \textbf{\bibinfo{volume}{85}},
  \bibinfo{pages}{1158} (\bibinfo{year}{2000}), \eprint{astro-ph/0003365},
  \urlprefix\url{http://dx.doi.org/10.1103/PhysRevLett.85.1158}.

\bibitem[{\citenamefont{Sikivie and Yang}(2009)}]{sikivie2009}
\bibinfo{author}{\bibfnamefont{P.}~\bibnamefont{Sikivie}} \bibnamefont{and}
  \bibinfo{author}{\bibfnamefont{Q.}~\bibnamefont{Yang}}
  (\bibinfo{year}{2009}), \eprint{0901.1106},
  \urlprefix\url{http://dx.doi.org/10.1103/PhysRevLett.103.111301}.

\bibitem[{\citenamefont{Silverman and Mallett}(2002)}]{silverman2002}
\bibinfo{author}{\bibfnamefont{M.~P.} \bibnamefont{Silverman}}
  \bibnamefont{and} \bibinfo{author}{\bibfnamefont{R.~L.}
  \bibnamefont{Mallett}}, \bibinfo{journal}{General Relativity and Gravitation}
  \textbf{\bibinfo{volume}{34}}, \bibinfo{pages}{633} (\bibinfo{year}{2002}),
  ISSN \bibinfo{issn}{00017701},
  \urlprefix\url{http://dx.doi.org/10.1023/A:1015934027224}.

\bibitem[{\citenamefont{Kain and Ling}(2010)}]{kain2010}
\bibinfo{author}{\bibfnamefont{B.}~\bibnamefont{Kain}} \bibnamefont{and}
  \bibinfo{author}{\bibfnamefont{H.~Y.} \bibnamefont{Ling}}
  (\bibinfo{year}{2010}), \eprint{1004.4692},
  \urlprefix\url{http://arxiv.org/abs/1004.4692}.

\bibitem[{\citenamefont{Arvanitaki and Dubovsky}(2010)}]{arvanitaki2010}
\bibinfo{author}{\bibfnamefont{A.}~\bibnamefont{Arvanitaki}} \bibnamefont{and}
  \bibinfo{author}{\bibfnamefont{S.}~\bibnamefont{Dubovsky}}
  (\bibinfo{year}{2010}), \eprint{1004.3558},
  \urlprefix\url{http://arxiv.org/abs/1004.3558}.

\bibitem[{\citenamefont{Rosa}(2010)}]{rosa2010}
\bibinfo{author}{\bibfnamefont{J.~G.} \bibnamefont{Rosa}}
  (\bibinfo{year}{2010}), \eprint{0912.1780},
  \urlprefix\url{http://arxiv.org/abs/0912.1780}.

\bibitem[{\citenamefont{Tegmark et~al.}(2006)\citenamefont{Tegmark, Aguirre,
  Rees, and Wilczek}}]{tegmark2006}
\bibinfo{author}{\bibfnamefont{M.}~\bibnamefont{Tegmark}},
  \bibinfo{author}{\bibfnamefont{A.}~\bibnamefont{Aguirre}},
  \bibinfo{author}{\bibfnamefont{M.~J.} \bibnamefont{Rees}}, \bibnamefont{and}
  \bibinfo{author}{\bibfnamefont{F.}~\bibnamefont{Wilczek}},
  \bibinfo{journal}{Physical Review D} \textbf{\bibinfo{volume}{73}},
  \bibinfo{pages}{023505+} (\bibinfo{year}{2006}),
  \urlprefix\url{http://dx.doi.org/10.1103/PhysRevD.73.023505}.

\bibitem[{\citenamefont{Mack}(2009)}]{mack2009a}
\bibinfo{author}{\bibfnamefont{K.~J.} \bibnamefont{Mack}}
  (\bibinfo{year}{2009}), \eprint{0911.0421},
  \urlprefix\url{http://arxiv.org/abs/0911.0421}.

\bibitem[{\citenamefont{Mack and Steinhardt}(2009)}]{mack2009b}
\bibinfo{author}{\bibfnamefont{K.~J.} \bibnamefont{Mack}} \bibnamefont{and}
  \bibinfo{author}{\bibfnamefont{P.~J.} \bibnamefont{Steinhardt}}
  (\bibinfo{year}{2009}), \eprint{0911.0418},
  \urlprefix\url{http://arxiv.org/abs/0911.0418}.

\bibitem[{\citenamefont{Linde and Noorbala}(2010)}]{linde2010}
\bibinfo{author}{\bibfnamefont{A.}~\bibnamefont{Linde}} \bibnamefont{and}
  \bibinfo{author}{\bibfnamefont{M.}~\bibnamefont{Noorbala}}
  (\bibinfo{year}{2010}), \eprint{1006.2170},
  \urlprefix\url{http://arxiv.org/abs/1006.2170}.

\bibitem[{\citenamefont{Hu}(1998)}]{hu1998b}
\bibinfo{author}{\bibfnamefont{W.}~\bibnamefont{Hu}}, \bibinfo{journal}{The
  Astrophysical Journal} \textbf{\bibinfo{volume}{506}}, \bibinfo{pages}{485}
  (\bibinfo{year}{1998}), \eprint{astro-ph/9801234},
  \urlprefix\url{http://dx.doi.org/10.1086/306274}.

\bibitem[{\citenamefont{Ma and Bertschinger}(1995)}]{bertschinger1995}
\bibinfo{author}{\bibfnamefont{C.-P.} \bibnamefont{Ma}} \bibnamefont{and}
  \bibinfo{author}{\bibfnamefont{E.}~\bibnamefont{Bertschinger}}
  (\bibinfo{year}{1995}), \eprint{astro-ph/9506072},
  \urlprefix\url{http://arxiv.org/abs/astro-ph/9506072}.

\bibitem[{\citenamefont{Dodelson}(2003)}]{book:dodelson}
\bibinfo{author}{\bibfnamefont{S.}~\bibnamefont{Dodelson}},
  \emph{\bibinfo{title}{Modern Cosmology}} (\bibinfo{publisher}{Academic
  Press}, \bibinfo{year}{2003}), \bibinfo{edition}{1st} ed., ISBN
  \bibinfo{isbn}{0122191412},
  \urlprefix\url{http://www.worldcat.org/isbn/0122191412}.

\bibitem[{\citenamefont{Peacock}(1998)}]{book:peacock}
\bibinfo{author}{\bibfnamefont{J.~A.} \bibnamefont{Peacock}},
  \emph{\bibinfo{title}{Cosmological Physics (Cambridge Astrophysics)}}
  (\bibinfo{publisher}{Cambridge University Press}, \bibinfo{year}{1998}),
  \bibinfo{edition}{1st} ed., ISBN \bibinfo{isbn}{0521422701},
  \urlprefix\url{http://www.worldcat.org/isbn/0521422701}.

\bibitem[{\citenamefont{Fox et~al.}(2004)\citenamefont{Fox, Pierce, and
  Thomas}}]{fox2004}
\bibinfo{author}{\bibfnamefont{P.}~\bibnamefont{Fox}},
  \bibinfo{author}{\bibfnamefont{A.}~\bibnamefont{Pierce}}, \bibnamefont{and}
  \bibinfo{author}{\bibfnamefont{S.}~\bibnamefont{Thomas}}
  (\bibinfo{year}{2004}), \eprint{hep-th/0409059},
  \urlprefix\url{http://arxiv.org/abs/hep-th/0409059}.

\bibitem[{\citenamefont{Perrotta and Baccigalupi}(1999)}]{perrotta1999}
\bibinfo{author}{\bibfnamefont{F.}~\bibnamefont{Perrotta}} \bibnamefont{and}
  \bibinfo{author}{\bibfnamefont{C.}~\bibnamefont{Baccigalupi}},
  \bibinfo{journal}{Physical Review D} \textbf{\bibinfo{volume}{59}},
  \bibinfo{pages}{123508+} (\bibinfo{year}{1999}),
  \urlprefix\url{http://dx.doi.org/10.1103/PhysRevD.59.123508}.

\bibitem[{\citenamefont{Bond et~al.}(1980)\citenamefont{Bond, Efstathiou, and
  Silk}}]{bond1980}
\bibinfo{author}{\bibfnamefont{J.~R.} \bibnamefont{Bond}},
  \bibinfo{author}{\bibfnamefont{G.}~\bibnamefont{Efstathiou}},
  \bibnamefont{and} \bibinfo{author}{\bibfnamefont{J.}~\bibnamefont{Silk}},
  \bibinfo{journal}{Physical Review Letters} \textbf{\bibinfo{volume}{45}},
  \bibinfo{pages}{1980} (\bibinfo{year}{1980}),
  \urlprefix\url{http://dx.doi.org/10.1103/PhysRevLett.45.1980}.

\bibitem[{\citenamefont{Amendola and Barbieri}(2005)}]{amendola2005}
\bibinfo{author}{\bibfnamefont{L.}~\bibnamefont{Amendola}} \bibnamefont{and}
  \bibinfo{author}{\bibfnamefont{R.}~\bibnamefont{Barbieri}}
  (\bibinfo{year}{2005}), \eprint{hep-ph/0509257},
  \urlprefix\url{http://dx.doi.org/10.1016/j.physletb.2006.08.069}.

\bibitem[{\citenamefont{Acharya et~al.}(2010)\citenamefont{Acharya, Bobkov, and
  Kumar}}]{acharya2010}
\bibinfo{author}{\bibfnamefont{B.~S.} \bibnamefont{Acharya}},
  \bibinfo{author}{\bibfnamefont{K.}~\bibnamefont{Bobkov}}, \bibnamefont{and}
  \bibinfo{author}{\bibfnamefont{P.}~\bibnamefont{Kumar}}
  (\bibinfo{year}{2010}), \eprint{1004.5138},
  \urlprefix\url{http://arxiv.org/abs/1004.5138}.

\end{thebibliography}

\end{document}